\documentclass{SciPost}
    
\hypersetup{
    colorlinks,
    linkcolor={red!50!black},
    citecolor={blue!50!black},
    urlcolor={blue!80!black}
}

\usepackage[bitstream-charter]{mathdesign}
\usepackage{mathtools} 
\DeclareSymbolFont{usualmathcal}{OMS}{cmsy}{m}{n}
\DeclareSymbolFontAlphabet{\mathcal}{usualmathcal}

\fancypagestyle{SPstyle}{
\fancyhf{}
\lhead{\colorbox{scipostblue}{\bf \color{white} ~SciPost Physics }}
\rhead{{\bf \color{scipostdeepblue} ~Submission }}

\fancyfoot[C]{\textbf{\thepage}}
}

\newcommand{\ket}[1]{\left\vert#1\right\rangle}
\newcommand{\bra}[1]{\left\langle#1\right\vert}

\def\Tr{\mbox{Tr}}

\begin{document}

\pagestyle{SPstyle}

\begin{center}{\Large \textbf{\color{scipostdeepblue}{
How compactness curbs entanglement growth in bosonic systems\\
}}}\end{center}

\begin{center}\textbf{
Stefan Aimet\textsuperscript{1$\star$},
Philipp Schmoll\textsuperscript{1},
Jens Eisert\textsuperscript{1,2},
J\"org Schmiedmayer\textsuperscript{3}\newline and
Spyros Sotiriadis\textsuperscript{4}
}\end{center}

\begin{center}
\begin{minipage}{0.85\textwidth}
\centering
{\bf 1} Freie Universität Berlin, Dahlem Center for Complex Quantum Systems, Department of Physics, 14195 Berlin, Germany\\
{\bf 2} Helmholtz-Zentrum Berlin für Materialien und Energie, \\ 14109 Berlin, Germany\\
{\bf 3} Vienna Center for Quantum Science and Technology, Atominstitut,\newline TU Wien, 1020 Vienna, Austria\\
{\bf 4} University of Crete, Department of Physics, 71003 Heraklion, Greece
\\[\baselineskip]
$\star$ \href{mailto:stefan.aimet@fu-berlin.de}{\small stefan.aimet@fu-berlin.de}
\end{minipage}
\end{center}

\section*{\color{scipostdeepblue}{Abstract}}
\textbf{\boldmath{%
Zero modes, here understood as degrees of freedom with vanishing confining frequency, play a central role in the nonequilibrium dynamics of bosonic systems. In Gaussian models, however, they drive an unbounded, logarithmic growth of entanglement entropy. We show that this divergence is not an intrinsic property of zero modes themselves, but arises specifically for non-compact zero modes. Their non-compact configuration space allows unbounded spreading in position space, while their continuous spectra enable indefinite dephasing in momentum space. By contrast, compact zero modes in compact bosonic systems behave fundamentally differently: Spreading and dephasing are eventually halted, so that compactness caps the entanglement entropy at a finite value, making its dynamical role most transparent in the presence of a zero mode. We demonstrate this mechanism in a minimal setting by comparing two coupled harmonic oscillators with two coupled quantum rotors. We then show that the same physics persists in many-body systems by contrasting an $N$-site compact rotor chain with the non-compact harmonic chain. Finally, we relate these insights to ultra-cold-atom realizations of compact quantum field theories. In particular, we clarify when a compact free-boson (Tomonaga–Luttinger liquid) description is required and when the commonly used non-compact massless Klein–Gordon model breaks down. Even when the initial state is accurately captured by a non-compact Gaussian description, compactness ultimately governs the late-time quench dynamics, curbing entanglement growth rather than allowing a dynamical divergence.
}}

\vspace{\baselineskip}



\vspace{10pt}
\noindent\rule{\textwidth}{1pt}
\tableofcontents
\noindent\rule{\textwidth}{1pt}
\vspace{10pt}

\section{Introduction}\label{section:Sec1}
The toolkit of quantum information theory~\cite{nielsen2010quantum} has fundamentally transformed how we characterize complex quantum systems using information-theoretic measures. Central among these is entanglement entropy~\cite{horodecki2009quantum}, which, in its various forms such as the von Neumann and Rényi entropies, provides a basis-independent quantification of quantum correlations in bipartite systems. Its importance is hard to overstate: Entanglement entropy is now a ubiquitous diagnostic across quantum information science~\cite{eisert2010colloquium}, condensed-matter physics~\cite{calabrese2005evolution,fagotti2008evolution}, and high-energy theory~\cite{ryu2006aspects}. In nonequilibrium settings it is a particularly powerful probe: Following a quantum quench—a sudden change of Hamiltonian parameters—its growth is routinely monitored to track information spreading, relaxation, and equilibration. Theoretically, entanglement dynamics after quenches have been explored extensively in both free and interacting systems, ranging from conformal field theories and integrable models~\cite{calabrese2005evolution,fagotti2008evolution,calabrese2007quantum} to lattice systems~\cite{audenaert2002entanglement,plenio2004dynamics,eisert2010colloquium,cotler2016entanglement}. Experimentally, rapid progress in ultra-cold atoms, trapped ions, and superconducting qubits has enabled measurements of entanglement dynamics in quantum many-body systems—most commonly through second-order Rényi entropies~\cite{islam2015measuring,kaufman2016quantum}, and more recently via protocols that access the von Neumann entropy itself~\cite{aimet2024experimentally}, which is the entanglement measure we use throughout this work.

\smallskip
A large portion of our theoretical understanding of entanglement dynamics in bosonic systems stems from studies of Gaussian models, where the Hamiltonian is quadratic and the relevant states are exponentials of quadratic forms. This focus is natural: Such systems are analytically tractable thanks to the machinery of Gaussian quantum information theory~\cite{weedbrook2012gaussian,eisert2003introduction}. From an experimental perspective, Gaussian states are also appealing because they can be prepared with high fidelity and reconstructed using tomographic schemes~\cite{gluza2020quantum,tajik2023verification,aimet2024experimentally}. Moreover, despite their simplicity, Gaussian models often capture the low-energy physics of interacting bosonic systems, which is frequently well described by an effective quadratic theory~\cite{giamarchi2003quantum}.
  
A paradigmatic example is the harmonic chain, whose quench dynamics have been studied extensively~\cite{audenaert2002entanglement,plenio2004dynamics,AnalyticalQuench,eisler2014entanglement,ghosh2018entanglement}. It remains invaluable for understanding entanglement growth, both as a discrete lattice system and, in the continuum limit, as the free scalar (Klein–Gordon) quantum field theory. However, Gaussian models such as the harmonic chain require particular care when zero modes are present, as these degrees of freedom play a distinguished role in both static and dynamical properties.
Throughout this work, we use the term \emph{mode} to denote a decoupled dynamical degree of freedom specified by a coordinate–momentum pair. A zero mode corresponds to such a degree of freedom with a vanishing confining potential, that is, a free-particle degree of freedom.\footnote{In other contexts, the term zero mode is sometimes used to refer directly to a zero-energy eigenstate. Since the Hamiltonian can be shifted by an arbitrary additive constant, absolute energy values are physically meaningless. Nevertheless, for definiteness we choose the Hamiltonians such that the lowest-energy eigenstate in the zero-mode sector is assigned zero energy.}

Already at the static level—for instance, in ground-state properties—the presence of such a zero mode leads to divergences in correlation functions and, correspondingly, in entanglement measures, necessitating explicit regularization or special treatment~\cite{mallayya2014zero,yazdi2017zero,chandran2019divergence,di2020entanglement,nenmeli2022when}. Dynamically, zero modes are especially important for long-time behavior. They arise most prominently in global frequency quenches that remove the confining on-site potential under Neumann or periodic boundary conditions. In Gaussian calculations, such quenches are found to induce a logarithmic growth of entanglement entropy at late times~\cite{hackl2018entanglement,chapman2019complexity,chandran2023dynamical,cotler2016entanglement,braccia2020complexity,jain2021log}.

\smallskip
This behavior is not merely of theoretical interest. A closely related situation arises in quantum field simulators based on ultra-cold Bose gases, where effective low-energy descriptions in terms of bosonic field theories can be engineered with high control. In particular, Ref.~\cite{aimet2024experimentally} realized a global quench from a low-temperature thermal state of a massive Klein–Gordon Hamiltonian to its massless counterpart and monitored the ensuing dynamics of the von Neumann entropy of a reduced subsystem, reporting that the zero mode spreads linearly in its target space over experimentally accessible timescales, a process commonly referred to as phase diffusion. This prompts a natural question: If such a quench could be implemented for arbitrarily long evolution times, would the logarithmic growth of this quantity predicted by the Gaussian theory persist indefinitely and thus be observable?

As we argue in this work, the answer is no. We recall that the effective low-energy description of the pre-quench Hamiltonian is given by the Klein–Gordon theory only as a quadratic, non-compact approximation to the underlying sine–Gordon field theory~\cite{giamarchi2003quantum, horvath2019nonequilibrium, castro2021branch,horvath2022inhomogeneous,szasz2024relaxation, szasz2024nonequilibrium, fitos2025quantum}, valid deep in the harmonic regime where a large on-site potential suppresses phase fluctuations. In this regime, the two theories describe the same initial state. However, their dynamics differ: The Klein–Gordon theory neglects the compactness of the field, which becomes relevant during time evolution. Consequently, the effective low-energy description governing the post-quench dynamics is not the non-compact massless Klein–Gordon field but the compact Tomonaga–Luttinger liquid~\cite{bastianello2020entanglement, bouchoule2025platforms}, the compact massless sine-Gordon field. While this distinction is often immaterial for short-time dynamics, it becomes decisive at long times, where the behavior is controlled by the dynamics of the zero mode.

In Gaussian (non-compact) systems, the zero mode is a free-particle degree of freedom supported on a non-compact configuration space. It therefore possesses a continuous spectrum (and hence a vanishing excitation gap), non-normalizable eigenstates, and allows unbounded spreading in position space together with indefinite dephasing in momentum space—precisely the mechanisms responsible for the logarithmic divergence of entanglement entropy observed in Gaussian quenches. By contrast, in compact theories the corresponding zero mode is compact: It is still a free-particle degree of freedom, but one defined on a bounded configuration space ($S^1$ rather than $\mathbb{R}$). As a result, its dynamics necessarily wrap around the compact domain rather than spreading without bound; the spectrum is discrete, eigenstates are normalizable, and dephasing ultimately saturates. Compactness therefore curbs entanglement growth and, in particular, removes the dynamical divergences that arise in Gaussian systems with a non-compact zero mode. While this mechanism is more general and applies even without a strict zero mode, we focus on zero-frequency quenches where such a mode emerges dynamically, as this setting provides the clearest illustration.

\begin{figure}[htpb]
\centering
\includegraphics[width=.9\linewidth]{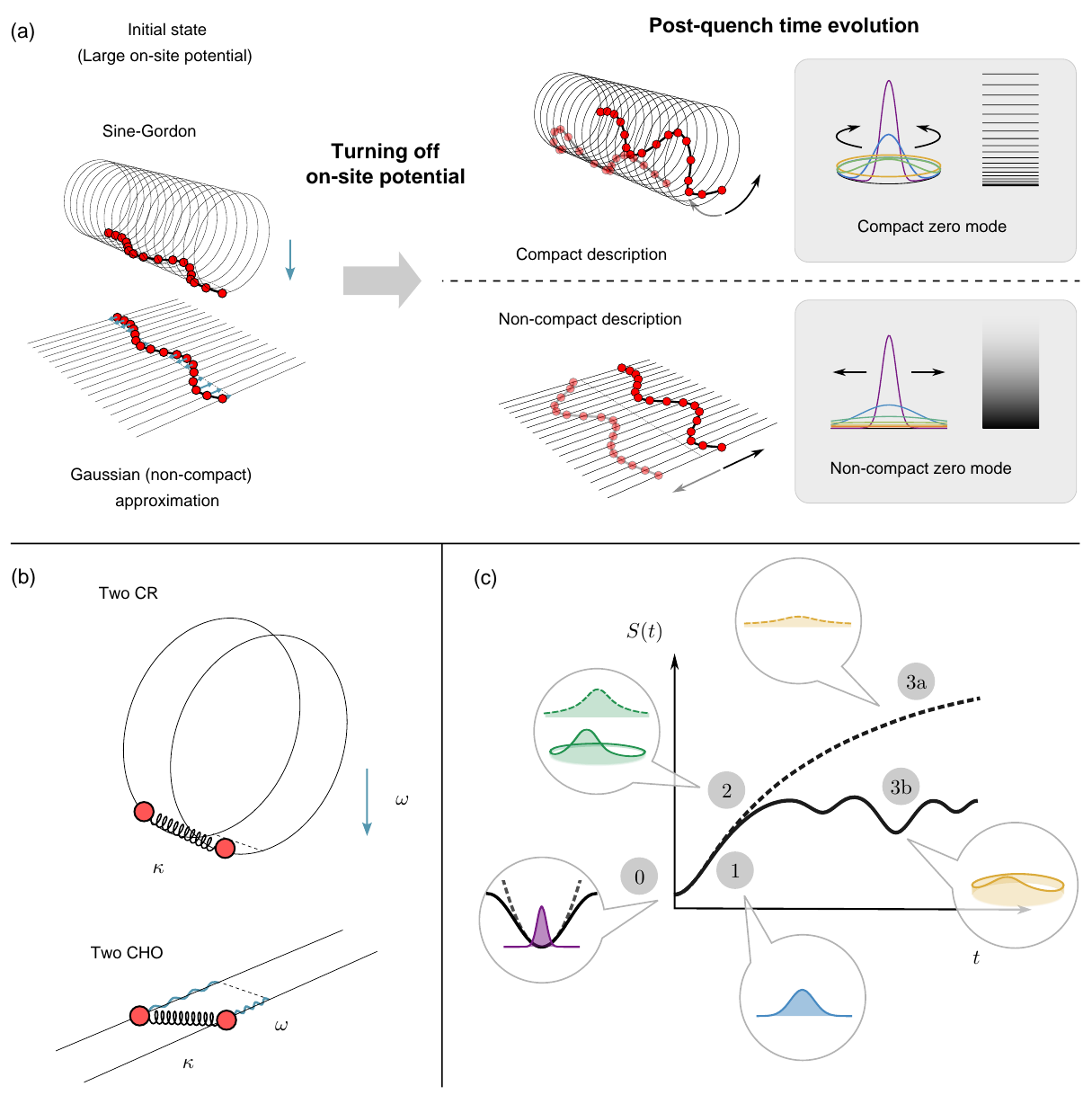}
 \caption{Schematic overview.
(a) Quantum field theory setting and quench protocol. Consider a system initially prepared in a low-temperature thermal state of the sine–Gordon quantum field theory on a finite spatial interval with Neumann boundary conditions. For a large on-site potential, phase fluctuations are suppressed and the initial state is well described by a quadratic, non-compact approximation—the massive Klein–Gordon theory. 
We then suddenly quench the on-site potential to zero, causing the center-of-mass to become a zero mode, which subsequently travels freely around the circle.
While the early-time dynamics are still captured by a non-compact description with an unbounded zero mode and continuous spectrum, the description of late-time behavior requires the compact theory, where the zero mode lives on a bounded domain and has a discrete spectrum.
(b) Minimal models. Two \emph{coupled quantum rotors} (CR, compact; see Sec.~\ref{section:Sec3}) and two \emph{coupled harmonic oscillators} (CHO, non-compact; see Sec.~\ref{section:Sec2}) provide the simplest realizations of the compact and non-compact theories in (a); their lattice generalizations form rotor and harmonic chains (see Sec.~\ref{section:Sec5}).
(c) Effect of compactness on entanglement growth. From the initial harmonic regime (0), the zero mode spreads in position space (equivalently dephases in momentum space) (1), where compact and non-compact dynamics coincide, until compactness sets in (2). Thereafter spreading continues, yielding logarithmic entanglement growth in the non-compact case (3a), but halts in the compact case, leading to saturation (3b).
}
  \label{fig:Fig1}
\end{figure}
We summarize the physical systems, quench protocol, and the mechanism by which compactness curbs entanglement growth schematically in Fig.~\ref{fig:Fig1}.
To make these ideas concrete, we study entanglement growth following global frequency quenches from the ground state of the pre-quench Hamiltonian to a post-quench Hamiltonian in which the on-site confining potential is set to zero, thereby generating a zero mode. To illustrate how compactness curbs entanglement growth, we proceed in stages of increasing complexity. We begin with a minimal two-site setting, comparing two coupled harmonic oscillators with their compact counterparts, two coupled quantum rotors, which already capture the essential physics of the mechanism. Such compact rotors are most familiarly encountered in kicked-rotor settings—canonical examples of quantum chaos~\cite{toloui2009quantum,pulikkottil2020entanglement,paul2020linear,santhanam2022quantum,sadia2022prethermalization}—a context only tangentially related to our aims here. 

We now outline how these ideas are concretely developed throughout this work. Section~\ref{section:Sec2} revisits the two-site harmonic-oscillator model and analyzes how the dynamical divergence of entanglement entropy arises. Section~\ref{section:Sec3} introduces the corresponding compact model of two coupled rotors. Section~\ref{section:Sec4} then presents a detailed comparison between the two models, combining numerical simulations with analytical estimates. Section~\ref{section:Sec5} extends these insights to many-body settings by studying short rotor chains—simulated using matrix product state methods—and contrasting them with harmonic chains. Finally, Section~\ref{section:Sec6} closes the loop by turning to continuum quantum field theories relevant for experimentally accessible ultra-cold-atom platforms. There we show how the compactness-driven mechanisms identified in the discrete models persist in the compact Tomonaga–Luttinger liquid but fail in the non-compact Klein–Gordon theory. We also outline practical subtleties and challenges that arise in the experimental analysis of these dynamics. Because interferometric measurements reconstruct the phase only modulo $2\pi$~\cite{schweigler2017experimental, schweigler2019correlations}, tracking the spreading of the zero mode on its compact domain becomes difficult once the phase distribution extends across the full $2\pi$ interval.

\section{Two coupled harmonic oscillators}\label{section:Sec2}
\subsection{Model}
We begin with a system of two \emph{coupled harmonic oscillators} (CHO), labeled by $n=1,2$. Each oscillator lives in the Hilbert space
\begin{equation}
\begin{aligned}
  \mathcal{H}^{{\rm CHO}}_{n}
    &= L^2(\mathbb{R})
      = \operatorname{span}\{\ket{x_n}\colon x_n\in\mathbb{R}\},
\end{aligned}
\end{equation}
so the full system occupies $\mathcal{H}^{\rm CHO}=\bigotimes_{n=1,2}\mathcal{H}^{{\rm CHO}}_{n}$.
The Hamiltonian of two coupled harmonic oscillators reads
\begin{equation}
    H^{\rm CHO}=\frac{1}{2}\left[p_1^2+p_2^2+\omega^2 (x_1^2+x_2^2)+\kappa(x_1-x_2)^2\right],
\end{equation}
where $\omega$ is the on-site potential frequency and $\kappa$ the coupling strength. The operators $x_n$ and $p_n$ are canonical position and momentum operators satisfying
$[x_n,p_m]=i\delta_{n,m}$.
Throughout this work, we set $\hbar=1$ unless stated otherwise. Introducing normal-mode coordinates,
\begin{equation}
x_{\pm}=\frac{x_1\pm x_2}{\sqrt{2}},\qquad p_{\pm}=\frac{p_1\pm p_2}{\sqrt{2}},
\end{equation}
preserves the commutators and decouples the Hamiltonian into two independent oscillators, $H^{\rm CHO}=H^{\rm CHO}_{+}+H^{\rm CHO}_{-}$, with
\begin{equation}
H^{\rm CHO}_{\pm}=\frac{1}{2}[p_{\pm}^2 + \omega_{\pm}^2 x_{\pm}^2]\,.
\end{equation}
The normal-mode frequencies are
\begin{equation}
    \omega_{+}=\omega,\qquad \omega_{-}=\sqrt{\omega^{2}+2\kappa},
\end{equation}
corresponding to center-of-mass and relative modes, respectively.

\subsection{Quench protocol}
We consider a global frequency quench in which the on-site confining potential is removed,
\begin{equation}
\omega_i=\omega \longrightarrow \omega_f=0,
\end{equation}
while keeping the coupling strength $\kappa$ fixed. Here and throughout, subscripts $i$ and $f$ denote pre- and post-quench quantities. The system is initialized in the ground state of the pre-quench Hamiltonian $H_i \coloneqq H^{\rm CHO}(\omega)$ and, for times $t>0$, evolves unitarily under the post-quench Hamiltonian $H_f \coloneqq  H^{\rm CHO}(\omega_f=0)$.

The quantity of interest is the bipartite entanglement between the two oscillators, quantified 
by the von Neumann entropy of the reduced density matrix of a single oscillator,
\begin{equation}
S_1(t) = -\mathrm{Tr}\!\left[\rho_1(t)\log \rho_1(t)\right],
\end{equation}
where $\rho_1(t)$ is obtained by tracing out oscillator~2 from the time-evolved state vector $\ket{\psi(t)}$.

\subsection{Spectrum of the post-quench Hamiltonian}
We now turn to the post-quench Hamiltonian $H_f^{\rm CHO}=H^{\rm CHO}_{f, +}+H^{\rm CHO}_{f, -}$ corresponding to the global frequency quench with $\omega_f=0$. The post-quench normal-mode frequencies are
\begin{equation}
\omega_{f,+} = 0, \qquad \omega_{f,-} = \sqrt{2\kappa}.
\end{equation}
Thus the center-of-mass mode becomes a zero mode, while the relative mode remains a standard bounded harmonic oscillator.
The relative-mode Hamiltonian $H^{\mathrm{CHO}}_{f,-}$ has the familiar discrete, evenly spaced spectrum
\begin{equation}
E_{k}^{(-)}=\omega_{f,-}\left(k+\frac{1}{2}\right),\qquad k=0,1,2,\hdots,
\end{equation}
with square-integrable Hermite–Gaussian eigenfunctions on $\mathbb{R}$. In contrast, the center-of-mass Hamiltonian $H^{\rm CHO}_{f,+}$ reduces, in the special case of the zero-frequency quench, to a free and unconfined degree of freedom on $\mathbb{R}$. Its energy spectrum is continuous,
\begin{equation}
E_{l}^{(+)}=\frac{l^2}{2},\qquad l\in\mathbb{R},
\end{equation}
and the corresponding eigenstates are plane waves $\propto e^{ilx_{+}}$, which are only delta-normalized rather than square-integrable.

\subsection{Time-evolved wave function}
The initial state vector $\ket{\psi^{\mathrm{CHO}}(0)}$ is the ground state of the pre-quench Hamiltonian $H^{\mathrm{CHO}}_i$.
In the normal-mode position basis it reads
\begin{equation}
\psi^{\mathrm{CHO}}(x_{+}, x_{-}; 0)
= \prod_{\nu=\pm} \left(\frac{\omega_{i,\nu}}{\pi}\right)^{1/4}
\exp\left[-\frac{1}{2}\omega_{i,\nu} x_{\nu}^{2}\right].
\end{equation}
For $t>0$, the state evolves as
\begin{equation}
\ket{\psi^{\mathrm{CHO}}(t)} = e^{-i H^{\mathrm{CHO}}_f t} \ket{\psi^{\mathrm{CHO}}(0)},
\end{equation}
and remains Gaussian at all times due to the quadratic form of the post-quench Hamiltonian $H^{\mathrm{CHO}}_f$.
Using the Lewis–Riesenfeld invariant method~\cite{lohe2008exact}, the wave function can be written (up to an overall phase) as
\begin{equation}\label{eq:time_evolved_state_real}
\psi^{\mathrm{CHO}}(x_{+}, x_{-}; t)
= \mathcal{N}
\exp\Bigg[-\frac{1}{2}\sum_{\nu=\pm}\big(A_\nu-iB_\nu\big) x_\nu^2
\Bigg],
\end{equation}
with
\begin{equation}
A_\nu(t) = \frac{\omega_{i,\nu}}{b_\nu^2}, \qquad
B_\nu(t) = \frac{\dot{b}_\nu}{b_\nu},\qquad \mathcal{N}(t)
= \prod_{\nu=\pm}
\left(\frac{A_{\nu}}{\pi }\right)^{1/4}\,.
\end{equation}
The scaling functions $b_\nu(\cdot)$ satisfy the non-linear Ermakov equation
\begin{equation}
\ddot{b}_{\nu}(t) + \omega_{f,\nu}^{2} b_{\nu}(t)
= \frac{\omega_{i,\nu}^{2}}{b_{\nu}^{3}(t)}\,,
\end{equation}
with $b_{\nu}(0)=1$ and $\dot{b}_{\nu}(0)=0$.
For the zero-frequency quench we obtain
\begin{equation}\label{eq:scaling functions_CHO}
\begin{aligned}
b_{+}(t)&=\sqrt{1+(\omega t)^2},\\
b_{-}(t)&=\sqrt{
\cos^2(\omega_{f,-}t)
+\Big(\tfrac{\omega_{i,-}}{\omega_{f,-}}\Big)^2\sin^2(\omega_{f,-}t)
}.
\end{aligned}
\end{equation}
Thus the center-of-mass mode exhibits unbounded growth characteristic of a non-compact zero mode, while the relative mode remains bounded and oscillatory.

\subsection{Reduced density matrix and entanglement entropy}
Expressing the time-evolved wave function in the original coordinates $(x_1,x_2)$, 
the reduced density matrix of oscillator~1 is obtained by tracing out site~2:
\begin{equation}
\rho^{\rm CHO}_1(x_1,x_1';t)=\int_{-\infty}^{\infty}dx_2\,\psi^{\rm CHO}(x_1,x_2;t)\psi^{\rm CHO,*}(x_1',x_2;t).
\end{equation}
Carrying out the Gaussian integral over $x_2$ yields
\begin{equation}
\begin{aligned}
\rho^{\rm CHO}_1(x_1,x_1';t)=\mathcal{N}_1\exp\left[-\frac{A}{2}(x_1^2+x_1'^2)+Bx_1x_1'+i\Phi(x_1^2-x_1'^2)\right],
\end{aligned}
\end{equation}
where
\begin{equation}
\begin{aligned}
    A&=\frac{2(A_+ + A_-)^2-[(A_+ - A_-)^2-(B_+ - B_-)^2]}{4(A_+ +  A_-)},\\
 B&=\frac{(A_+-A_-)^2+(B_+-B_-)^2}{4(A_++A_-)},\\
\Phi &= \frac{B_++B_-}{4}-\frac{(A_+-A_-)(B_+-B_-)}{4(A_++A_-)}.
\end{aligned}
\end{equation}
The normalization factor is given by
\begin{equation}
\mathcal{N}_1=|\mathcal{N}|^2\sqrt{\frac{2\pi}{A_+ + A_-}}\,.
\end{equation}
Since the local phase $i\Phi(x_1^2-x_1'^2)$ does not affect the eigenvalue spectrum of $\rho^{\rm CHO}_1$, 
it can be omitted when computing the entanglement entropy. 
The resulting kernel is of the standard Mehler form, which can be diagonalized analytically using Mehler’s formula for Hermite functions~\cite{srednicki1993entropy,ghosh2018entanglement}. 
The eigenvalues are
\begin{equation}
    \lambda_k=(1-\xi)\xi^k, \qquad k=0,1,2,\dots,
\end{equation}
with
\begin{equation}\label{eq:xi}
    \xi=\frac{\chi}{1+\sqrt{1-\chi^2}}, \qquad \chi(t)=\frac{B(t)}{A(t)}\,.
\end{equation}
The von Neumann entanglement entropy hence becomes
\begin{equation}
    S_1^{\rm CHO}(t)=-\sum_{k=0}^{\infty}\lambda_k\log{\lambda_k}
    =-\log{(1-\xi)}-\frac{\xi}{1-\xi}\log{(\xi)}\,.
\end{equation}

\subsection{Late-time entanglement growth}
To extract the late-time behavior of the entanglement entropy, we analyze the asymptotics of the Gaussian kernel parameters entering the reduced density matrix $\rho^{\rm CHO}_1(x_1,x_1';t)$. Using the large-$t$ form of the scaling functions $b_{\nu}(\cdot)$ from Eq.~\eqref{eq:scaling functions_CHO}, the normal-mode coefficients behave as
\begin{equation}
\begin{aligned}
A_+(t)\sim \frac{1}{t^2},\qquad B_+(t)\sim \frac{1}{t},\qquad A_-(t),\, B_-(t)=\mathcal{O}(1),\qquad (t\to\infty),
\end{aligned}
\end{equation}
where $\mathcal{O}(1)$ denotes bounded, oscillatory functions of time.
Substituting these asymptotic forms into the exact expressions for $A(t)$ and $B(t)$, and hence $\chi(t)$, $\xi(t)$, and $S_1^{\rm CHO}(t)$, we obtain
\begin{equation}
S_1^{\rm CHO}(t)\sim \log t + \mathcal{O}(1),\qquad (t\to\infty),
\end{equation}
where the bounded $\mathcal{O}(1)$ contribution arises from the relative mode. The logarithmic growth is, therefore, directly driven by the zero mode.

\medskip
This asymptotic behavior is fully consistent with the rigorous classification of entanglement growth in quadratic Hamiltonians developed in Ref.~\cite{hackl2018entanglement}. There, three dynamical sectors are identified:
\begin{itemize}
\item \emph{Stable sector:} The Hamiltonian is bounded below and diagonalizable on a complete orthonormal basis, implying that the entanglement entropy remains bounded for all time.
\item \emph{Metastable sector:} The Hamiltonian is still bounded below but possesses non-normalizable eigenvectors, leading to a continuous spectrum and logarithmic entanglement growth.
\item \emph{Unstable sector:} The Hamiltonian is unbounded from below, producing exponential growth of mode amplitudes and linear-in-time entanglement growth.
\end{itemize}
In the present CHO quench, these sectors correspond directly to the post-quench frequencies $\omega_{f,\pm}$: Positive, zero, and imaginary frequencies lead to stable, metastable, and unstable behavior, respectively (see also Ref.~\cite{chandran2023dynamical}). Restricting to real, non-negative $\omega_f\ge 0$, only the first two cases are relevant. When $\omega_f>0$, both modes remain stable, ensuring bounded entanglement entropy. For the quench to $\omega_f=0$, as considered here, however, the center-of-mass frequency $\omega_{f,+}$ vanishes, producing a free, unbounded zero mode on $\mathbb{R}$ and placing the system in the metastable sector—hence the logarithmic growth derived above.

\subsection{Position-space interpretation}
While the preceding analysis, together with the results of Ref.~\cite{hackl2018entanglement}, establishes the logarithmic growth of entanglement entropy on rigorous grounds, it is useful to develop a more intuitive position-space interpretation of this dynamical divergence—an interpretation that will later allow for a clean contrast with the compact coupled-rotor model.

\medskip
Starting from the Gaussian time-evolved probability density for each mode $\nu=\pm$,
\begin{equation}
|\psi^{(\nu)}(x_\nu;t)|^2=\Big(\frac{A_\nu}{\pi}\Big)^{1/2}e^{-A_\nu x_\nu^2},
\end{equation}
the variances and widths follow as
\begin{equation}
    \langle x_\nu^2(t)\rangle = \frac{1}{2A_\nu(t)},
    \qquad
    \sigma_{x_\nu}(t)=\frac{1}{\sqrt{2A_\nu(t)}}\,.
\end{equation}
Using the large-$t$ asymptotics of $A_\nu(t)$, one finds
\begin{equation}
\sigma_{x_{+}}(t)\sim t,
\qquad
\sigma_{x_{-}}(t)=\mathcal{O}(1),\qquad (t\to\infty),
\end{equation}
reflecting unbounded spreading of the zero mode and bounded oscillatory motion of the relative mode. Consequently, the on-site width $\sigma_{x_1}(t)$ inherits a linear growth in $t$ up to bounded oscillatory corrections.

\medskip
To expose the mechanism behind the divergence of the entanglement entropy, it is instructive to examine the reduced density matrix directly. Writing it in symmetric and anti-symmetric coordinates,
\begin{equation}
X_{\mathrm{s}}=\frac{x_1+x_1'}{\sqrt{2}},\qquad 
X_{\mathrm{a}}=\frac{x_1-x_1'}{\sqrt{2}},
\end{equation}
one finds the Gaussian kernel (up to an irrelevant phase)
\begin{equation}
\begin{aligned}
\rho^{\rm CHO}_1(X_{\mathrm{s}},X_{\mathrm{a}};t)=\mathcal{N}_1\exp\left[
-\frac{X_{\mathrm{s}}^2}{2\ell_{X_{\mathrm{s}}}^2} - \frac{X_{\mathrm{a}}^2}{2\ell_{X_{\mathrm{a}}}^2}\right],
\end{aligned}
\end{equation}
with coherence lengths
\begin{equation}
\ell_{X_{\mathrm{s}}}=(A-B)^{-1/2},\qquad \ell_{X_{\mathrm{a}}}=(A+B)^{-1/2}\,.
\end{equation}
It is useful to express the entanglement parameter $\xi$ directly in terms of these coherence lengths.
Starting from Eq.~\ref{eq:xi} for $\xi$ in terms of $\chi$, one finds
\begin{equation}\label{eq:xi_coherence_lengths}
\xi
=\frac{\ell_{X_{\mathrm s}}-\ell_{X_{\mathrm a}}}
{\ell_{X_{\mathrm s}}+\ell_{X_{\mathrm a}}}.
\end{equation}
Using the large-$t$ asymptotics of $A_\pm$ and $B_\pm$, the coherence lengths behave as
\begin{equation}
\ell_{X_{\mathrm{s}}}(t)\sim t, \qquad \ell_{X_{\mathrm{a}}}(t)=\mathcal{O}(1),\qquad (t\to\infty).
\end{equation}
As a consequence, Eq.~\eqref{eq:xi_coherence_lengths} yields
\begin{equation}
\xi(t)\sim 1-\mathcal{O}\!\left(\frac{1}{t}\right),
\qquad (t\to\infty),
\end{equation}
which directly implies a logarithmic growth of the entanglement entropy at late times.

The mechanism behind the divergence of the entanglement entropy can now be seen directly.
The reduced density matrix becomes increasingly extended along the symmetric direction
$X_{\mathrm{s}}$ as a result of the unbounded spreading of the non-compact zero mode, while
remaining bounded along the anti-symmetric direction $X_{\mathrm{a}}$.
This anisotropic growth is controlled by the decay of the zero-mode coefficient
$A_+(t)\sim t^{-2}$, which causes the coherence length $\ell_{X_{\mathrm{s}}}$ to diverge,
whereas $\ell_{X_{\mathrm{a}}}$ stays finite.
Because the spreading takes place on the non-compact configuration space $\mathbb{R}$, it
never saturates: The reduced density matrix acquires support over an ever-growing region in
the $(x_1,x_1')$ plane, producing an increasingly broad spectrum of Schmidt coefficients and,
consequently, a logarithmically diverging entanglement entropy.
In position space, the origin of the divergence is therefore the unbounded spreading of the non-compact zero mode.

\subsection{Momentum-space interpretation}
The time-evolved momentum-space wave function follows from the two-dimensional Fourier transform of $\psi^{\mathrm{CHO}}(x_+,x_-;t)$. Using the normal-mode coordinates, it is given by
\begin{equation}
\tilde{\psi}^{\mathrm{CHO}}(p_+,p_-;t)
=\frac{1}{2\pi}\!\int_{-\infty}^{\infty}\!dx_+dx_-\, 
e^{-i(p_+x_+ + p_-x_-)}
\psi^{\mathrm{CHO}}(x_+,x_-;t).
\end{equation}
Substituting the Gaussian form of Eq.~\eqref{eq:time_evolved_state_real}, each integral can be computed analytically, yielding (up to an overall phase) a wave function of identical Gaussian form
\begin{equation}
\tilde{\psi}^{\mathrm{CHO}}(p_+,p_-;t)=\tilde{\mathcal{N}}\exp\Big[-\frac{1}{2}\sum_{\nu=\pm}(\tilde{A}_{\nu}-i\tilde{B}_{\nu})p_{\nu}^2\Big]
\end{equation}
with
\begin{equation}
\tilde{A}_{\nu}(t)=\frac{A_{\nu}}{A_{\nu}^2+B_{\nu}^2},\qquad \tilde{B}_{\nu}(t)=-\frac{B_{\nu}}{A_{\nu}^2+B_{\nu}^2},\qquad \tilde{\mathcal{N}}(t)=\prod_{\nu=\pm}\Big(\frac{\tilde{A}_{\nu}}{\pi}\Big)^{1/4}.
\end{equation}
The momentum widths follow directly from the Gaussian and mirror the position-space structure.

For the zero-frequency quench, the center-of-mass mode has a time-independent real coefficient
\begin{equation}
    \tilde{A}_+(t)=\frac{1}{\omega},
\end{equation}
and therefore a constant momentum width,
\begin{equation}
\sigma_{p_+}(t)=\sqrt{\frac{1}{2\tilde{A}_+}}=\sqrt{\frac{\omega}{2}}=\mathrm{const.}
\end{equation}
This is exactly what one expects: In the post-quench Hamiltonian the zero mode becomes a free particle, and since $[H_{f,+},p_+]=0$, its momentum distribution must be conserved. The relative mode remains a bounded oscillator, so $\tilde{A}_-(t)$ stays oscillatory and $\sigma_{p_-}(t)=\mathcal{O}(1)$; accordingly, the on-site momentum width $\sigma_{p_1}(t)$ also remains bounded.

\medskip
From these considerations alone, nothing in momentum space appears to diverge, raising the question of how the entanglement entropy can nevertheless grow without bound. The resolution is that, while the real coefficient $\tilde{A}_+$ controlling the momentum width remains constant, the zero-mode phase coefficient develops a linearly growing contribution,
\begin{equation}
    \tilde{B}_+(t)= -t,\qquad (t\to\infty),
\end{equation}
which induces increasingly strong dephasing among momentum components.

Consequently, the reduced density matrix in momentum space,
written in symmetric/anti-symmetric variables,
\begin{equation}
P_{\mathrm{s}}=\frac{p_1+p_1'}{\sqrt{2}},\qquad
P_{\mathrm{a}}=\frac{p_1-p_1'}{\sqrt{2}},
\end{equation}
takes the form
\begin{equation}
\tilde{\rho}^{\rm CHO}_1(P_{\mathrm{s}},P_{\mathrm{a}};t)
=\tilde{\mathcal{N}}_1\,
\exp\!\left[
-\frac{P_{\mathrm{s}}^2}{2\ell_{P_{\mathrm{s}}}^2(t)}
-\frac{P_{\mathrm{a}}^2}{2\ell_{P_{\mathrm{a}}}^2(t)}
\right],
\end{equation}
with momentum-space coherence lengths
\begin{equation}
\ell_{P_{\mathrm{s}}}(t)=(\tilde{A}-\tilde{B})^{-1/2},\qquad
\ell_{P_{\mathrm{a}}}(t)=(\tilde{A}+\tilde{B})^{-1/2},
\end{equation}
and composite coefficients
\begin{equation}
\begin{aligned}
\tilde{A}&=\frac{2(\tilde{A}_+ + \tilde{A}_-)^2 - [(\tilde{A}_+ - \tilde{A}_-)^2 - (\tilde{B}_+ - \tilde{B}_-)^2]}{4(\tilde{A}_+ + \tilde{A}_-)},\\
\tilde{B}&=\frac{(\tilde{A}_+ - \tilde{A}_-)^2 + (\tilde{B}_+ - \tilde{B}_-)^2}{4(\tilde{A}_+ + \tilde{A}_-)}.
\end{aligned}
\end{equation}
The normalization factor is 
\begin{equation}
\tilde{\mathcal{N}}_1(t)=|\tilde{\mathcal{N}}|^2\sqrt{2\pi/(\tilde{A}_+ + \tilde{A}_-)}.    
\end{equation}
Substituting the explicit forms of $\tilde{A}_{\pm}(t)$ and $\tilde{B}_{\pm}(t)$ yields the asymptotic behavior
\begin{equation}
\ell_{P_{\mathrm{s}}}(t)=\text{const.},\qquad
\ell_{P_{\mathrm{a}}}(t)\sim \frac{1}{t},\qquad (t\to\infty).
\end{equation}
As in position space, it is useful to express the entanglement parameter in terms of these coherence
lengths.
Defining $\tilde{\xi}=\tilde{B}/\tilde{A}$, one finds
\begin{equation}\label{eq:xi_momentum_coherence_lengths}
\tilde{\xi}
=\frac{\ell_{P_{\mathrm{s}}}-\ell_{P_{\mathrm{a}}}}
{\ell_{P_{\mathrm{s}}}+\ell_{P_{\mathrm{a}}}}.
\end{equation}
Since the Schmidt spectrum is representation-independent, $\tilde{\xi}=\xi$, as expected.
Using the asymptotic behavior above, Eq.~\eqref{eq:xi_momentum_coherence_lengths} yields
\begin{equation}
\tilde{\xi}(t)\sim 1-\mathcal{O}\!\left(\frac{1}{t}\right),
\qquad (t\to\infty),
\end{equation}
reproducing the same late-time divergence obtained in position space.

The physical origin of the divergent entanglement growth is thus equally transparent in momentum
space.
Although the momentum distribution of the zero mode remains fixed, corresponding to a finite width
along $P_{\mathrm{s}}$, the zero-mode phase coefficient $\tilde{B}_+(t)\sim -t$ grows linearly in time and induces
increasingly strong dephasing along the
anti-symmetric direction $P_{\mathrm{a}}$. As a result, the reduced density matrix becomes progressively more diagonal in momentum space. Because the zero-mode spectrum is continuous, this dephasing never saturates, providing the complementary momentum-space interpretation of the same logarithmic entanglement growth.

\section{Two coupled rotors}\label{section:Sec3}
\subsection{Model}
We now turn to the compact counterpart of the coupled harmonic oscillators: Two \emph{coupled rotors} (CR), each representing an angular degree of freedom confined to a circle. The angular coordinates $x_n\in[-\pi,\pi)$ for $n=1,2$ are defined modulo $2\pi$, i.e., $x_n\equiv x_n+2\pi$ with conjugate angular momenta $p_n$. Strictly speaking, $x_n$ is a phase variable~\cite{Carruthers1968}, so canonical commutation relations
between $x_n$ and $p_n$ are not well defined.
Instead, the algebra is formulated in terms of the unitary operators $e^{- i x_n}$, which satisfy
$[e^{-i x_n},p_m]=\delta_{n,m}e^{-i x_n}$.
Each rotor lives in the Hilbert space
\begin{equation}
\begin{aligned}
  \mathcal{H}^{{\rm CR}}_n
    &= L^2(S^1)
      = \operatorname{span}\{\ket{x_n}\colon x_n \in[-\pi,\pi)\},
\end{aligned}
\end{equation} 
so that the two-rotor system occupies
$\mathcal{H}^{\rm CR}=\bigotimes_{n=1,2}\mathcal{H}^{{\rm CR}}_n\simeq L^2\bigl(T^2\bigr)$, where $T^2=S^1\times S^1$ is the two-torus.
The Hamiltonian of two coupled rotors reads 
\begin{equation}\label{eq:rotor_hamiltonian}
\begin{aligned}
 H^{\rm CR}=\frac{1}{2}\left(p_1^2+p_2^2\right)+\omega^2\left(2 -\cos{(x_1)} -\cos{(x_2)}\right)+\kappa \left(1-\cos{(x_1-x_2)}\right).
\end{aligned}
\end{equation}
In the small-angle limit, where the compact topology of $S^1$ is not yet visible, expanding 
\begin{equation}
\cos{(x)}= 1-\tfrac{1}{2}x^2+O(x^4)
\end{equation}
reduces Eq.~\eqref{eq:rotor_hamiltonian} to the Hamiltonian of two coupled harmonic oscillators discussed in Sec.~\ref{section:Sec2}, and the algebra reduces correspondingly to canonical commutation relations. In this sense, the CHO model is recovered as the Gaussian small-angle limit of the compact rotor model.

\subsection{Spectrum of the post-quench Hamiltonian}
In the CHO model on $\mathbb{R}$, removing the confining potential turns the center-of-mass mode into a free, non-compact zero mode with a continuous spectrum, which is directly responsible for the logarithmic entanglement growth. The coupled-rotor model is qualitatively different: Here, each coordinate is compactified on $S^1$, and the Hamiltonian~\eqref{eq:rotor_hamiltonian} is invariant under discrete $2\pi$ shifts
\begin{equation}
T_n:x_n\mapsto x_n+2\pi,\quad n=1,2.
\end{equation}
Compactness ensures that the spectrum of $H^{\rm CR}$ is purely discrete and bounded below for all values of $\omega$ and $\kappa$ (see the spectral theory of Schr\"odinger operators on compact manifolds, e.g., Ref.~\cite{reed1978iv}).
This discreteness is already visible in the ``free'' kinetic term: On the circle, the angular
momentum operator $p_n$ has a discrete spectrum, with eigenstates $\ket{p_n}$ labeled by
integers $p_n\in\mathbb{Z}$ (we use the same symbol for this operator and its eigenvalues throughout).
One thus has
\begin{equation}
p_n\ket{p_n}=p_n\ket{p_n},\qquad p_n\in\mathbb{Z},
\end{equation}
with properly normalized plane waves
\begin{equation}
    \langle x_n|p_n\rangle =\frac{1}{\sqrt{2\pi}}e^{ip_n x_n},
\end{equation}
which are square-integrable on $[-\pi,\pi)$.

\medskip
For the zero-frequency quench $\omega_f=0$, the post-quench rotor Hamiltonian again separates at the operator level into center-of-mass and relative parts, $H^{\rm CR}_f=H^{\rm CR}_{f,+}+H^{\rm CR}_{f,-}$, where
\begin{equation}
    H^{\rm CR}_{f,+}=\frac{1}{2}p_+^2,\qquad H^{\rm CR}_{f,-}=\frac{1}{2}p_-^2+\kappa\left(1-\cos{(\sqrt{2}x_-)}\right).
\end{equation}
The center-of-mass sector $H^{\rm CR}_{f,+}$ now describes a compact zero mode. Its spectrum is discrete,
\begin{equation}
    E_{p}^{(+)}=\frac{p^2}{4}, \qquad p\in\mathbb{Z},
\end{equation}
with properly normalizable plane-wave eigenstates $\psi_{p}^{(+)}(x_+)$ on a bounded configuration space. Thus, in stark contrast to the CHO case, the zero mode no longer lives on $\mathbb{R}$ and therefore cannot spread indefinitely in real space. The relative sector $H^{\rm CR}_{f,-}$ instead yields the familiar Mathieu spectrum~\cite{DLMF}. Its eigenvalues are
\begin{equation}
    E_{q,\sigma}^{(-)}=\kappa+\frac{1}{4}\begin{cases}
a_q(-2\kappa), & \sigma = a,\qquad q = 0,1,2,\dots,\\
b_q(-2\kappa), & \sigma = b,\qquad q = 1,2,3,\dots,
    \end{cases}
\end{equation}
where $a_q(\cdot)$ and $b_q(\cdot)$ are the even and odd Mathieu characteristic values, respectively. The corresponding (unnormalized) eigenfunctions in the relative coordinate are
\begin{equation}
\psi_{q,\sigma}^{(-)}(x_-)=
\begin{cases}
{\rm ce}_q\left(\dfrac{x_-}{\sqrt{2}},-2\kappa\right), & \sigma = a,\\
{\rm se}_q\left(\dfrac{x_-}{\sqrt{2}},-2\kappa\right), & \sigma = b,
\end{cases}
\end{equation}
where ${\rm ce}_q$ and ${\rm se}_q$ are the even and odd Mathieu functions.

\medskip
There is, however, an important subtlety. The original coordinates $(x_1,x_2)$ live on the
torus $T^2$, so the full wave function must be $2\pi$-periodic in each variable separately. This single-valuedness constraint ties the center-of-mass and relative sectors together and prevents the Hilbert space from factorizing as $\mathcal{H}_+\otimes\mathcal{H}_-$. Instead, it decomposes as
\begin{equation}
\mathcal{H}^{\rm CR} \;\simeq\; (\mathcal{H}_+^{\rm CR, \,even}\otimes\mathcal{H}_-^{\rm CR, \, even})
\;\oplus\;
(\mathcal{H}_+^{\rm CR, \, odd}\otimes\mathcal{H}_-^{\rm CR, \,odd}),
\end{equation}
where $\mathcal{H}_\pm^{\rm CR, \, even/odd}$ denote the subspaces spanned by eigenstates of the
post-quench Hamiltonians $H^{\rm CR}_{f,\pm}$ with even or odd integer momentum quantum
numbers, respectively. In particular, the allowed combinations of center-of-mass and
relative quantum numbers obey a global parity constraint,
\begin{equation}
  p - q \equiv 0 \quad (\mathrm{mod}\, 2),
\end{equation}
so that the spectrum naturally decomposes into even and odd $p$-sectors. As a result, the compact zero mode does not live on an independent copy of $S^1$, but on a one-dimensional projection of the torus whose admissible eigenstates are correlated with those of the relative sector through this parity condition.

\subsection{Momentum-space truncation and numerical implementation}
The natural basis for the coupled-rotor Hamiltonian $H^{\rm CR}$ is the angular-momentum (Fourier) basis
\begin{equation}
     \{\ket{p_1,p_2}:p_n\ket{p_1,p_2}=p_n \ket{p_1,p_2},\qquad p_n\in\mathbb{Z}\}\,,
\end{equation}
in which the compactness of the configuration space is encoded in the discreteness of the angular momenta.
Although compactness guarantees a discrete spectrum, the Hilbert space remains infinite-dimensional, since each $p_n$ ranges over all integers. For numerical calculations we therefore truncate the angular momenta to a finite window
\begin{equation}
p_n\in\{-M,\hdots, M\}\,,
\end{equation}
giving a finite total Hilbert space dimension of $(2M+1)^2$. In this truncated basis, the Hamiltonian matrix elements are
\begin{equation}
\begin{aligned}
\langle p_1,p_2|H^{\rm CR}|p'_1,p'_2\rangle
&=\frac{1}{2}(p_1^2 + p_2^2)\,\delta_{p_1,p'_1}\,\delta_{p_2,p'_2}
\\[4pt]
&\quad+\;\omega^2\Bigl[2\,\delta_{p_1,p'_1}\,\delta_{p_2,p'_2}
-\tfrac12\bigl(\delta_{p_1-1,p'_1}+\delta_{p_1+1,p'_1}\bigr)\,\delta_{p_2,p'_2}
-\tfrac12\,\delta_{p_1,p'_1}\,\bigl(\delta_{p_2-1,p'_2}+\delta_{p_2+1,p'_2}\bigr)\Bigr]
\\[4pt]
&\quad+\;\kappa\Bigl[\delta_{p_1,p'_1}\,\delta_{p_2,p'_2}
-\tfrac12\bigl(\delta_{p_1-1,p'_1}\,\delta_{p_2+1,p'_2}
+\delta_{p_1+1,p'_1}\,\delta_{p_2-1,p'_2}\bigr)\Bigr].
\end{aligned}
\end{equation}
Here, the first term is the kinetic energy, the second term is the on-site potential, and the third term describes the nearest-neighbor coupling in momentum space. Since the full rotor Hamiltonian $H^{\rm CR}$ in Eq.~\eqref{eq:rotor_hamiltonian} is not analytically diagonalizable, this truncated momentum-space representation provides the natural starting point for numerical calculations.

\medskip
A key virtue of the rotor model is that momentum-space truncation is exceptionally well controlled. This ultimately rests on rigorous analytic bounds showing that Fourier coefficients of eigenfunctions of analytic periodic potentials decay exponentially fast~\cite{reed1978iv}. As a consequence, all physically relevant states—including the pre-quench ground state and its subsequent time evolution—are strongly concentrated in a finite low-momentum window. High-momentum components are exponentially suppressed, and observables converge rapidly as the cutoff $M$ is increased. 

\begin{figure}[htpb]
    \centering
    \includegraphics[width=\linewidth]{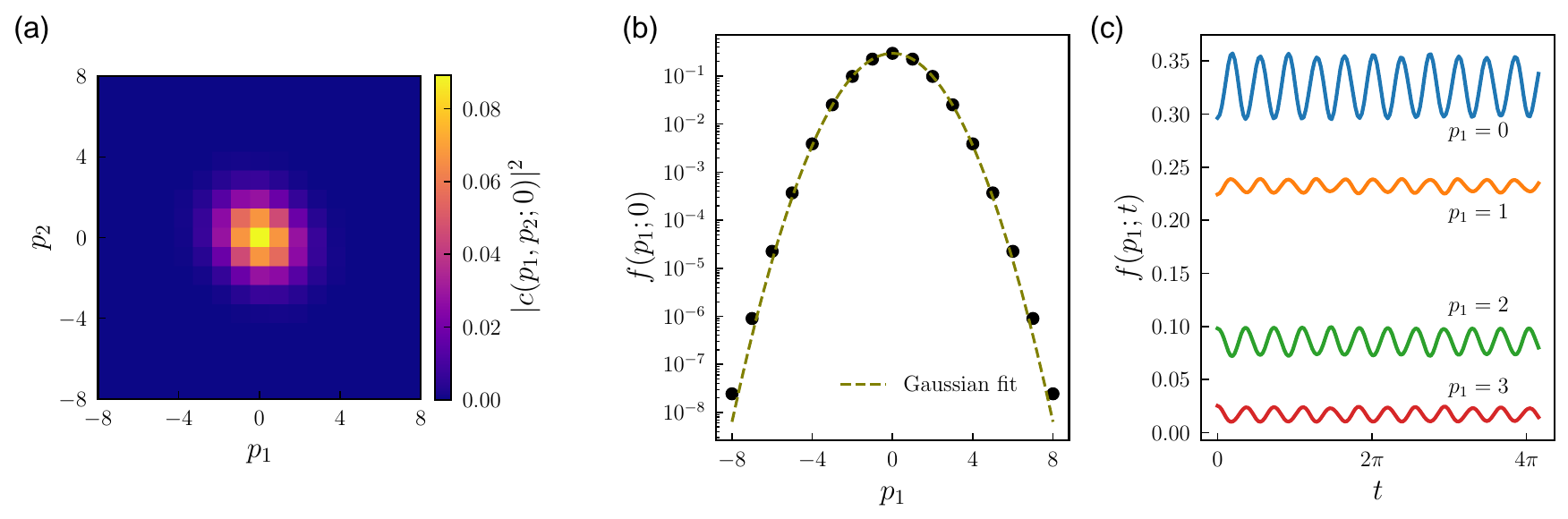}
    \caption{Momentum-space structure of the pre-quench ground state in the two-site \emph{coupled rotor} (CR) model.
(a) Joint probability density $|c(p_1,p_2;0)|^2$ in the truncated angular-momentum basis $p_n\in\{-M,\ldots,M\}$, showing strong localization around $(p_1,p_2)=(0,0)$.
(b) Marginal distribution $f(p_1;0)$ on a semi-logarithmic scale together with a Gaussian fit, demonstrating rapid decay of high-momentum components.
(c) Time evolution of selected momentum occupations $f(p_1;t)$, showing that localization in momentum space is preserved under the post-quench dynamics.
Parameters: $\omega^2=5$, $\kappa=10$.}
    \label{fig:Fig2}
\end{figure}

\noindent Numerical analysis in Fig.~\ref{fig:Fig2} shows that the Fourier weights
\begin{equation}
 |c(p_1,p_2;0)|^2=|\langle p_1,p_2|\psi^{\rm CR}(0)\rangle|^2   
\end{equation}
of the initial state are tightly localized in the $(p_1,p_2)$ plane, and—crucially—the time-evolved coefficients $c(p_1,p_2;t)$ remain confined to essentially the same low-energy sector. The intuition from the CHO carries over: In the zero-frequency quench one again has $[H^{\rm CR}_{f,+},p_+]=0$, so the center-of-mass momentum distribution should not broaden. The only subtlety is that the initial coupled-rotor ground state does not strictly factorize into $+$ and $-$ sectors, due to the presence of the on-site potential, and is, therefore not an exact eigenstate of $p_+$. Nevertheless, its support in $(p_+,p_-)$ space is sharply localized, and hence so is its distribution in $(p_1,p_2)$. 

By extracting the momentum-space marginal
\begin{equation}
f(p_1;0)=\sum_{p_2}|c(p_1,p_2;0)|^2,
\end{equation}
we indeed find an excellent fit to an exponentially decaying profile, as illustrated in Fig.~\ref{fig:Fig2}(b). The subsequent time evolution of the occupations $f(p_1;t)$, shown in Fig.~\ref{fig:Fig2}(c), confirms that this strong localization in momentum space persists under the post-quench dynamics.

\section{Comparing dynamics of two coupled harmonic oscillators and rotors}\label{section:Sec4}
The goal of this section is to understand how compactness modifies the unbounded entanglement growth that follows a zero-frequency quench. To this end we compare, in detail, the dynamics of two coupled rotors with those of the corresponding non-compact coupled harmonic oscillators.

\begin{figure}[htpb]
\centering
\includegraphics[width=.9\linewidth]{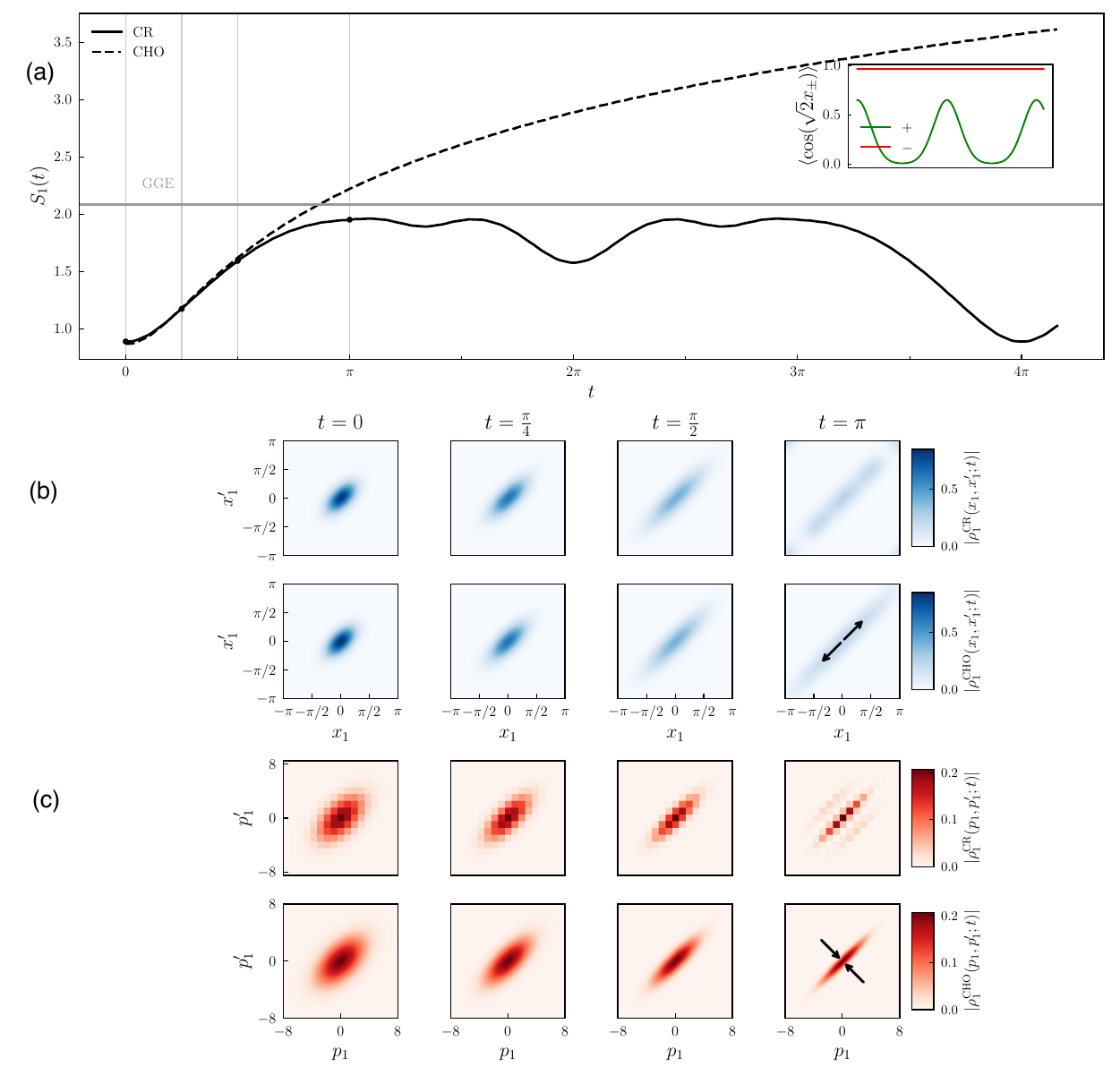}
\caption{Dynamics after a zero-frequency quench in the two-site 
\emph{coupled-rotor} (CR) model.
(a) Single-site entanglement entropy $S_1(t)$. At early times the CR dynamics coincide with those of the non-compact \emph{coupled harmonic oscillator} (CHO), before compactness becomes relevant. While the CHO entropy grows logarithmically, the CR entropy saturates at a finite value, remaining below the \emph{generalized Gibbs ensemble} (GGE) prediction, and exhibits oscillations. The inset shows $\langle\cos(\sqrt{2}x_{\pm})\rangle$, revealing collapse–revival dynamics of the center-of-mass mode $(+)$ and an effectively frozen relative mode $(-)$.
(b) Magnitude of the reduced single-site density matrix in position space, $|\rho^{\rm CR}_1(x_1,x_1';t)|$, at selected times.
(c) Magnitude of the reduced single-site density matrix in 
momentum space, $|\rho^{\rm CR}_1(p_1,p_1';t)|$, at the same times as in (b).
Parameters: $\kappa=100$, $\omega^2=10$.}
    \label{fig:Fig3}
\end{figure}

\subsection{Choice of parameter regime}
We focus on a parameter regime in which the CR ground state and the early dynamics are, to excellent approximation, identical to those of the CHO, while compactness becomes relevant only at late times. Concretely, we consider
\begin{equation}
2\kappa \gg \omega^2 \gg 1\,.
\end{equation}
The large pre-quench frequency $\omega^2 \gg 1$ ensures that the CR ground state is tightly localized near the origin within the fundamental domain, so that the cosine potentials are well approximated by their quadratic expansion. As a result, the CR and CHO ground states become practically indistinguishable.

At the same time, the strong-coupling condition 
$2\kappa \gg \omega^2$ produces a sharp separation of scales between the normal modes. We choose this regime specifically so that the relative mode is essentially frozen out by the quench. This is quantified by the ratio
\begin{equation}
    r_-:=\frac{\omega_{i,-}}{\omega_{f,-}}
    =\frac{\sqrt{\omega^2+2\kappa}}{\sqrt{2\kappa}}
    =\sqrt{1+\frac{\omega^2}{2\kappa}}\,.
\end{equation}
Requiring $r_-\approx 1$ implies $\omega^2/(2\kappa)\ll 1$, and in turn the CHO scaling function satisfies $b_-(t)\approx 1$ for all times (cf.\ Eq.~\eqref{eq:scaling functions_CHO}). Semi-classically, the deep cosine well traps the relative coordinate into tiny, high-frequency oscillations around its minimum, keeping this mode effectively frozen. This realizes a Born–Oppenheimer–type hierarchy: Just as in molecular physics, where heavy nuclei move slowly while light electrons adjust almost instantaneously, the small ratio $\omega^2/(2\kappa)\ll 1$ enforces a similar separation of timescales, making the center-of-mass mode the slow degree of freedom and the relative mode the fast one.
Consequently, the post-quench evolution is dominated by the center-of-mass mode and cleanly isolates the zero-mode contribution to the entanglement dynamics.

\subsection{Late-time dynamics}
At early times, the compact center-of-mass coordinate behaves indistinguishably from a free particle
on the real line, as it has not yet “felt’’ the compact topology. Accordingly, the CR and CHO dynamics coincide initially. This agreement inevitably breaks down once the spread of the center-of-mass wave packet becomes
comparable to the size of its compact domain (see Fig.~\ref{fig:Fig3}(a)). Beyond this point, the CR entanglement entropy no longer follows the logarithmic growth of the CHO. Instead, it levels off, capped by the compact two-torus geometry, while the CHO entropy continues its
unbounded metastable rise. The CR therefore reaches a finite maximum,
\begin{equation}
S^{\rm CR}_{1,\max}=\max_t S_1^{\rm CR}(t),
\end{equation}
after which it remains bounded.

\medskip
A particularly transparent diagnostic of the dynamics of the two normal modes is provided by the
expectation values
\begin{equation}
\langle \cos{(\sqrt{2}x_{\pm})}\rangle(t),
\end{equation}
shown in the inset of Fig.~\ref{fig:Fig3}(a).
The relative mode remains essentially frozen in this parameter regime.
The center-of-mass mode, by contrast, exhibits a collapse--revival pattern.
Starting from a sharply localized state, $\langle\cos(\sqrt{2}\,x_+)\rangle(t)$ decays as the wave
packet spreads and eventually wraps around its compact domain.
At this point the center-of-mass mode becomes maximally delocalized, and the reduced single-site
density matrix—shown in position space in Fig.~\ref{fig:Fig3}(b) and in momentum space in
Fig.~\ref{fig:Fig3}(c)—is maximally mixed.
Correspondingly, the entanglement entropy reaches $S^{\rm CR}_{1,\max}$.

Beyond this regime, clear discrepancies between the CR and CHO dynamics emerge.
For the CHO, the coherence length $\ell_{X_{\mathrm{s}}}$ in position space continues to grow without
bound, while in momentum space the coherence length $\ell_{P_{\mathrm{a}}}$ keeps shrinking, as
discussed earlier. In the compact rotor model, by contrast, this mechanism of indefinite spreading and dephasing is
halted by compactness.
This is reflected directly in the reduced density matrix and, consequently, in the entanglement
entropy, which no longer grows but instead exhibits a sequence of (quasi-)recurrences.
For the present parameter regime, revivals of the single-site entanglement entropy
$S_1^{\rm CR}(t)$ associated with the zero mode occur at integer multiples of $4\pi$.

Due to inversion symmetry, signatures of these revivals already appear at half the full revival time.
The latter can be estimated analytically in the zero-mode--dominated regime from the time-evolved
wave function.
Since the relative mode remains effectively frozen, its time dependence may be neglected to leading
order.
Expanding the center-of-mass sector in eigenstates of the compact zero-mode Hamiltonian yields
\begin{equation}
\psi^{\rm CR}(x_+,x_-;t)
\approx
\left(\sum_{p} 
c_p^{(+)}\, e^{-i p^2 t /4}\,\psi^{(+)}_p(x_+)\right)
\psi^{(-)}(x_-;0).
\end{equation}
The recurrence structure follows directly from the quadratic phase factors $e^{-i p^2 t/4}$, whose
partial and full rephasing generate quasi-recurrences and revivals, respectively. The corresponding full revival time is therefore approximately given by integer multiples of
$8\pi$.

\subsection{How compactness curbs late-time entanglement growth}
The physical mechanism behind the finite ceiling $S^{\rm CR}_{1,\max}$ can be understood most
clearly by comparison with the CHO discussed in Sec.~\ref{section:Sec2}.
There, the logarithmic divergence of the entanglement entropy had two complementary origins: unbounded spreading of the non-compact zero mode in position space and indefinite dephasing
associated with its continuous momentum spectrum.
Equivalently, the symmetric position-space coherence length $\ell_{X_{\mathrm{s}}}$ grows linearly in
time, while the anti-symmetric momentum-space coherence length $\ell_{P_{\mathrm{a}}}$ shrinks as
$1/t$.

For the coupled rotors the same intuition applies, but compactness changes the late-time behavior
qualitatively.
In position space, the center-of-mass mode is now compact: Its support is confined to a compact domain, so $\ell_{X_{\rm s}}$ can grow only until the wave packet has wrapped around its compact domain. Beyond this point further spreading is impossible and $\ell_{X_{\rm s}}(\cdot)$ saturates. In momentum space, the center-of-mass momentum takes integer values; the resulting discrete level spacing prevents indefinitely increasing dephasing, and $\ell_{P_{\rm a}}(\cdot)$ approaches a finite limit.

In both representations the conclusion is the same: Compactness removes the possibility of unbounded spreading in configuration space and of indefinitely increasing dephasing in a continuous spectrum. The entanglement entropy, therefore,  cannot diverge logarithmically but instead grows only up to a finite ceiling $S^{\rm CR}_{1,\max}$.

A natural question is how this intuitive picture is reflected in the reduced density matrix
corresponding to $S^{\rm CR}_{1,\max}$. Numerical analysis (Appendix~\ref{app:estimate}) shows that the intuition is broadly correct: The diagonal ensemble—obtained by discarding
all relative phases between post-quench eigenstates while retaining their initial
populations—already provides a reasonable estimate of $S^{\rm CR}_{1,\max}$. A slightly tighter estimate is obtained from the block-diagonal ensemble, which preserves coherences
between degenerate eigenstates within the symmetry-protected even and odd parity sectors
imposed by the global parity constraint. Small but noticeable deviations persist due to
residual dynamical coherences, but these are quantitatively minor. For completeness,
Appendix~\ref{app:bound} also provides a (fairly loose) uniform-in-time upper bound on the
entanglement entropy, thereby establishing rigorously that $S_1^{\rm CR}(\cdot)$ cannot
diverge. 

\subsection{Analytic estimate of $S^{\rm CR}_{1,\rm max}$}
As seen above, the entanglement entropy of the coupled rotors never exceeds a finite ceiling $S^{\rm CR}_{1,\max}$. In this subsection we provide a quantitative estimate of this ceiling.
Rather than using the diagonal or block-diagonal ensembles—which capture the late-time dephasing picture well—we turn here to the \emph{generalized Gibbs ensemble} (GGE)~\cite{eisert2015quantum,gogolin2016equilibration,Polkovnikov,CalabreseEsslerFagotti11,gluza2019equilibration,AnalyticalQuench}. The GGE has two advantages. First, it provides an excellent quantitative estimate of $S^{\rm CR}_{1,\max}$ in our two-rotor setting (see Appendix~\ref{app:estimate}). Second, it is the natural object for many-body and field-theoretic quenches—the context ultimately relevant for the comparison with experiment.

The GGE maximizes the von Neumann entropy subject to all nontrivial conserved quantities of the post-quench dynamics. In our model, the only conserved quantities are the energies of the decoupled center-of-mass and relative sectors. The GGE density matrix is therefore
\begin{equation}
  \rho^{\rm CR}_{\rm GGE}
  =\frac{1}{Z}\exp\Bigl(-\lambda_{+}H^{\rm CR}_{f,+} - \lambda_{-}H^{\rm CR}_{f,-}\Bigr),
\end{equation}
with partition function
\begin{equation}
Z=\Tr\!\left[e^{-\lambda_{+}H^{\rm CR}_{f,+} - \lambda_{-}H^{\rm CR}_{f,-}}\right].
\end{equation}
The Lagrange multipliers $\lambda_{\pm}$ are fixed by matching the conserved post-quench energies,
\begin{equation}
\Tr(\rho^{\rm CR}_{\rm GGE}H^{\rm CR}_{f,\pm}) = E_{\pm},
\qquad
E_{\pm} = \langle \psi^{\rm CR}(0)|H^{\rm CR}_{f,\pm}|\psi^{\rm CR}(0)\rangle.
\end{equation}

Because of the single-valuedness constraint on the torus, the joint eigenstates $\ket{p,q,\sigma}$ of $H^{\rm CR}_{f,+}$ and $H^{\rm CR}_{f,-}$ must satisfy $p-q\equiv 0\;(\mathrm{mod}\;2)$, so the GGE becomes
\begin{equation}
\rho^{\rm CR}_{\rm GGE}
= \frac{1}{Z}
\sum_{p,q,\sigma}\Pi_{p,q}\,
e^{-\lambda_{+}E_{p}^{(+)} - \lambda_{-}E_{q,\sigma}^{(-)}}
\ket{p,q,\sigma}\bra{p,q,\sigma},
\qquad
\Pi_{p,q}=\frac12\!\left[1+(-1)^{p-q}\right].
\end{equation}
Correspondingly,
\begin{equation}
Z(\lambda_{+},\lambda_{-})
= Z_{+}^{\rm even}(\lambda_{+})Z_{-}^{\rm even}(\lambda_{-})
 + Z_{+}^{\rm odd}(\lambda_{+})Z_{-}^{\rm odd}(\lambda_{-}),
\end{equation}
with
\begin{equation}
Z_{+}^{\rm even/odd}(\lambda_{+})
= \sum_{\substack{p\in\mathbb{Z}\\ p\;\mathrm{even/odd}}}
e^{-\lambda_{+}E_{p}^{(+)}},
\qquad
Z_{-}^{\rm even/odd}(\lambda_{-})
= \sum_{\sigma}\sum_{\substack{q\in\mathbb{Z}\\ q\;\mathrm{even/odd}}}
e^{-\lambda_{-}E_{q,\sigma}^{(-)}}.
\end{equation}
In our chosen regime, the relative mode remains frozen in its deep ground state $(q=0,\sigma=a)$ throughout the dynamics. The parity constraint then enforces even total momentum $p\in 2\mathbb{Z}$, and the GGE reduces to
\begin{equation}
\rho^{\rm CR}_{\rm GGE}
\approx
\left(\sum_{p\in2\mathbb{Z}} w_p\ket{p}_+\bra{p}_+\right)
\otimes
\ket{\psi^{(-)}(0)}\bra{\psi^{(-)}(0)},
\qquad
w_p=\frac{e^{-\lambda_{+}p^2/4}}{\sum_{r\in2\mathbb{Z}}e^{-\lambda_{+}r^2/4}}.
\end{equation}
In the deep-well limit, $\ket{\psi^{(-)}(0)}$ is well approximated by the Gaussian ground state vector of the corresponding harmonic oscillator (cf.\ Sec.~\ref{section:Sec2}). Writing it in the relative-momentum basis $\{\ket{d}_-:d\in2\mathbb{Z}\}$,
\begin{equation}
\ket{\psi^{(-)}(0)}
= \sum_{d\in2\mathbb{Z}} c_d\ket{d}_-,
\qquad
|c_d|^2
= \frac{\exp(-d^2/2\omega_{i,-})}{\sum_{s\in2\mathbb{Z}}\exp(-s^2/2\omega_{i,-})}.
\end{equation}
Converting to single-rotor momenta using $p_1=(p+d)/2$, $p_2=(p-d)/2$ and tracing out rotor~2 yields a diagonal reduced GGE density matrix
\begin{equation}
\rho^{\rm CR}_{1,{\rm GGE}}
= \sum_{m\in\mathbb{Z}} \lambda_m\ket{m}\bra{m},
\qquad
\lambda_m=\sum_{p\in2\mathbb{Z}} w_p\,|c_{2m-p}|^2.
\end{equation}
Both $w_p$ and $|c_{2m-p}|^2$ are sharply peaked Gaussians, so we approximate the sums by integrals over $q=p/2$ and obtain
\begin{equation}
\lambda_m
\simeq
\frac{1}{\sqrt{2\pi\sigma_m^2}}
\exp\!\left[-\frac{m^2}{2\sigma_m^2}\right],
\qquad
\sigma_m^2 = \frac{1}{2\lambda_{+}} + \frac{\omega_{i,-}}{4}.
\end{equation}
Treating $m$ as continuous, the single-site GGE entropy becomes
\begin{equation}
S^{\rm CR}_{1,{\rm GGE}}
= -\sum_m\lambda_m\ln\lambda_m
\simeq \frac12\ln\!\left(2\pi e\,\sigma_m^2\right).
\end{equation}
To relate $\lambda_{+}$ to the conserved center-of-mass energy $E_{+}$ we evaluate
\begin{equation}
Z_{+}^{\rm even}(\lambda_{+})
\simeq \int_{-\infty}^{\infty}\frac{dp}{2}\,e^{-\lambda_{+}p^2/4}
= \sqrt{\frac{\pi}{\lambda_{+}}},
\end{equation}
so that
\begin{equation}
E_{+}
= -\partial_{\lambda_{+}}\ln Z_{+}^{\rm even}
\simeq \frac{1}{2\lambda_{+}}.
\end{equation}
Thus
\begin{equation}
\sigma_m^2
\simeq E_{+} + \frac{\omega_{i,-}}{4},
\end{equation}
and therefore
\begin{equation}
S^{\rm CR}_{1,{\rm GGE}}
\simeq \frac12\ln\!\left[
2\pi e\left(E_{+} + \frac{\omega_{i,-}}{4}\right)
\right].
\end{equation}
Using the Gaussian CHO estimate $E_{+}\simeq \omega/4$ gives the compact expression
\begin{equation}\label{eq:SGGE estimate}
S^{\rm CR}_{1,{\rm GGE}}
\simeq \frac12
\ln\!\left[
\frac{\pi e}{2}\,\bigl(\omega+\sqrt{\omega^2+2\kappa}\,\bigr)
\right],
\end{equation}
which provides our analytic estimate of the entropy ceiling $S^{\rm CR}_{1,\max}$.

\subsection{Further parameter regimes}
\begin{figure}[htpb]
\centering
\includegraphics[width=\linewidth]{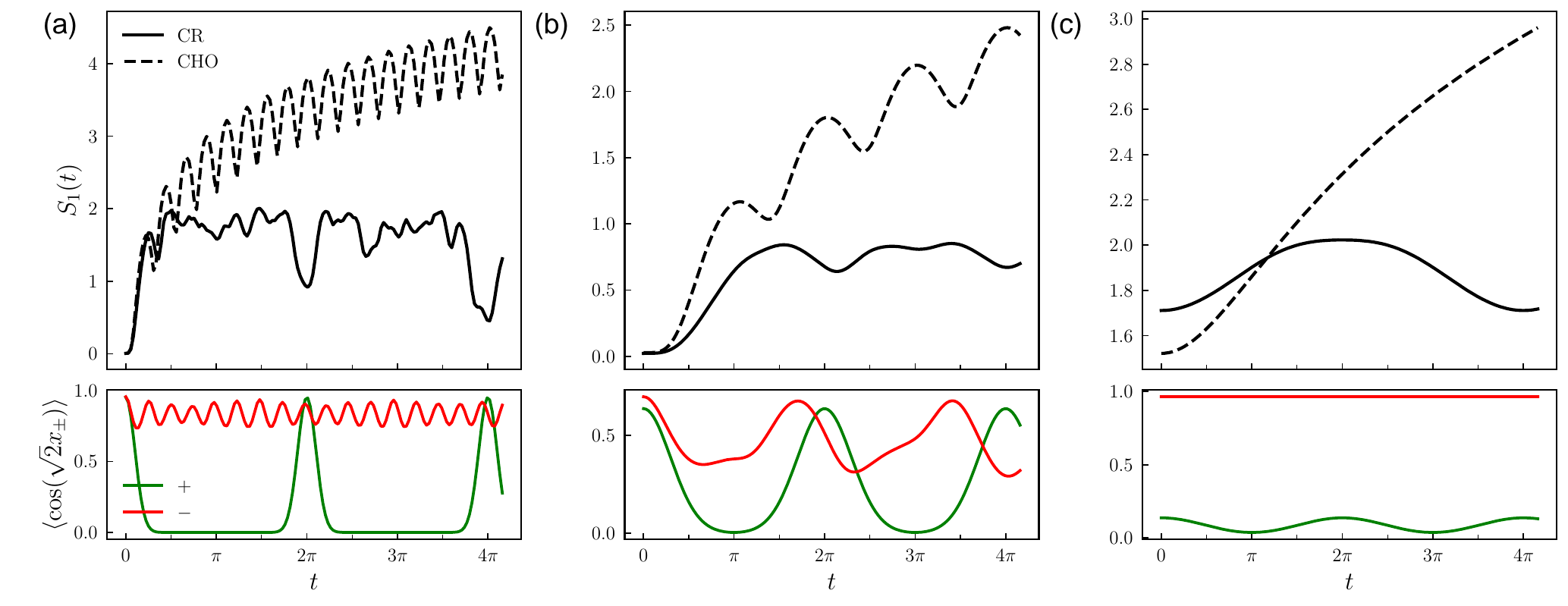}
   \caption{Dynamics after a zero-frequency quench in the two-site \emph{coupled-rotor} (CR) model for different parameter regimes.
Top panels: Single-site entanglement entropy $S_1(t)$ for the CR (solid) and the corresponding \emph{coupled harmonic oscillators} (CHO, dashed). 
Bottom panels: Expectation values $\langle\cos(\sqrt{2}x_{\pm})\rangle$.
(a) Regime with $\omega^2 \gg 2\kappa \gg 1$ (here $\omega^2 = 100$, $\kappa = 10$), where the initial state and the early-time dynamics are still well described by the Gaussian harmonic approximation, but the relative mode is no longer frozen.
(b) Intermediate regime with $\omega^2 \sim \kappa \sim 1$ (here $\omega^2 = 1.5$, $\kappa = 0.5$), where the initial state remains approximately Gaussian but the relative mode contributes substantially; these parameters match those used in Fig.~\ref{fig:Fig5}.
(c) Regime with a very weak trap and strong coupling (here $\omega^2 = 0.1$, $\kappa = 100$), where compactness is already visible in the initial state.
}
    \label{fig:Fig4}
\end{figure}

So far we have focused on the regime $2\kappa \gg \omega^2 \gg 1$, where the initial CR state is extremely well approximated by the CHO ground state and the relative mode is effectively frozen. In this limit the early-time CR dynamics coincide with those of the CHO, and the clean logarithmic entanglement growth of the CHO is driven solely by its non-compact zero mode. Figure~\ref{fig:Fig4} illustrates how the behavior changes for other parameter regimes.

In Fig.~\ref{fig:Fig4}(a) we consider a regime with $\omega^2\gg2\kappa\gg1$.
Here the initial state, as well as the early-time dynamics, are still well described by the Gaussian harmonic approximation. However, in contrast to the regime considered previously, the relative mode is no longer frozen out and now contributes to the dynamics, giving rise to pronounced oscillatory deviations about the logarithmic CHO growth. Nonetheless, compactness remains irrelevant at early times, and only at later times does it induce saturation of the entanglement entropy.

Fig.~\ref{fig:Fig4}(b) shows a more generic case with $\kappa\sim\omega^2\sim1$, matching the parameters used in the rotor-chain simulations of Sec.~\ref{section:Sec5}. The initial state is still approximately Gaussian, but the relative mode now matters substantially. As a result, $S_1^{\rm CR}(t)$ departs from the CHO behavior already at very early times and exhibits high-frequency oscillations from the relative sector. The approach to $S^{\rm CR}_{1,\max}$ is broadened, and the revival structure is shifted and distorted compared to the clean multiples of $4\pi$ observed when only the center-of-mass mode dominates.

Finally, Fig.~\ref{fig:Fig4}(c) illustrates a regime with $\omega^2\ll1$ and $\kappa\gg1$. The trap is now so weak that the initial state already senses the periodicity of the cosine potential and is visibly non-Gaussian at $t=0$. For sufficiently large $\kappa$, however, the relative motion remains effectively frozen, and the dynamics are again governed primarily by the compact center-of-mass sector. In this case $S_1^{\rm CR}(t)$ reaches its maximum at $t=2\pi$ and subsequently exhibits smooth oscillations around its ceiling.

\section{Dynamics of coupled harmonic oscillator and rotor chains}\label{section:Sec5}
Having analyzed the two-site models as pedagogical examples that expose the role of compact zero modes and the limitations of the harmonic approximation, we now turn to the many-body setting with $N$ degrees of freedom. We consider an $N$-site chain, bipartition it symmetrically into two halves of length $N_A = N/2$, and study the time evolution of the half-chain entanglement entropy $S_{N/2}(.)$ after a global frequency quench.

In direct analogy with the two-site problem, we compare two models:
an $N$-site harmonic chain and an $N$-site coupled-rotor chain. The former provides a non-compact Gaussian reference, in which a zero mode yields unbounded logarithmic entanglement growth. The latter is the compact counterpart, where the same quench produces a compact zero mode whose dynamics curb entanglement growth.

\subsection{Harmonic chain with Neumann boundary conditions}
For the harmonic chain, one typically imposes one of three standard \emph{boundary conditions}  (BCs): Dirichlet BCs, which fix the endpoints ($x_{0}=x_{N+1}=0$); periodic BCs, which identify them ($x_{n+N}=x_{n}$); or Neumann BCs, which leave the boundaries free ($x_{0}=x_{1}$, $x_{N+1}=x_{N}$).

Under a quench to $\omega_f=0$, Dirichlet BCs are the only choice that do not produce a zero mode, while periodic and Neumann BCs each do. Since our interest lies precisely in the late-time behavior driven by the zero mode, and because an open chain with free endpoints matches the experimental setting (see Sec.~\ref{section:Sec6}) best, we focus on Neumann BCs.

The Hamiltonian of an $N$-site chain of coupled harmonic oscillators with Neumann BCs is
\begin{equation}
    \begin{aligned}
	H_{(N)}^{\rm CHO}
 =\frac{1}{2}\left[\sum_{n=1}^{N}(p_n^2+\omega^2x_n^2)
 +\kappa\sum_{n=1}^{N-1}(x_{n}-x_{n+1})^2\right].
    \end{aligned}
\end{equation}
As in the two-site case, one may diagonalize $H^{\rm CHO}_{(N)}$ in terms of normal modes and solve the post-quench dynamics using the associated Ermakov equations~\cite{ghosh2018entanglement,chandran2023dynamical}. For Neumann BCs, the spatially uniform center-of-mass mode $X_{+}=\frac{1}{\sqrt{N}}\sum_{n=1}^{N}x_n$ (with conjugate momentum $P_+$) becomes a non-compact zero mode with post-quench Hamiltonian $P_{+}^{2}/2$ and continuous spectrum, and is again responsible for the unbounded logarithmic entanglement growth at the many-body level.

Although the wave function-based normal-mode approach yields fully analytical expressions, the covariance-matrix formalism~\cite{weedbrook2012gaussian,audenaert2002entanglement,di2019entanglement} is numerically more robust at large $N$. Because the state remains Gaussian at all times, the time evolution of the full many-body state is determined completely by the linear Heisenberg evolution of $(x_n(.),p_n(.))$, or equivalently by the evolution of the covariance matrix. The half-chain entanglement entropy is then obtained from the symplectic spectrum of the reduced covariance matrix.

\subsection{Coupled-rotor chain and tensor-network methods}
The compact counterpart of the harmonic chain is the $N$-site chain of coupled rotors with Neumann BCs, described by the Hamiltonian
\begin{equation}
\begin{aligned}
    H^{\rm CR}_{(N)}
    =\frac{1}{2}\Bigg[\sum_{n=1}^{N}p_n^2 
    + 2\omega^2\sum_{n=1}^{N}\bigl(1-\cos x_n\bigr)
    + 2\kappa\sum_{n=1}^{N-1}\bigl(1-\cos(x_n-x_{n+1})\bigr)\Bigg].
\end{aligned}
\end{equation}
Each angle operator $x_n$ is compact, $x_n \equiv x_n + 2\pi$, and the conjugate angular momenta $p_n$ have integer spectra, $p_n\ket{p_n} = p_n \ket{p_n}$ with $p_n\in\mathbb{Z}$. Compactness guarantees that the many-body spectrum is purely discrete, but the local Hilbert space remains infinite-dimensional. As a result, exact diagonalization becomes impractical already at modest system sizes $N\gtrsim 4$.

To access larger chains than achievable with exact diagonalization, we employ one-di\-men\-sional \emph{tensor network} (TN) methods~\cite{Orus2014,Orus2019,Cirac2021,eisert2010colloquium}. In particular, we represent $H^{\rm CR}_{(N)}$ efficiently as a \emph{matrix product operator} (MPO). Because $H^{\rm CR}_{(N)}$ contains only on-site terms and nearest-neighbor couplings, its MPO representation requires only a modest bond dimension. The many-body ground state of the pre-quench Hamiltonian $H^{\rm CR}_{(N)}(\omega)$ is then obtained variationally as a \emph{matrix product state} (MPS) using a variant of a \emph{density matrix renormalization group} (DMRG) algorithm~\cite{White92,Oestlund1995,Dukelsky1998,PerezGarcia2007,Schollwoeck2011}, after introducing a controlled truncation of the local angular-momentum basis. Real-time dynamics after the quench to $\omega_f=0$ are simulated using the \emph{time-dependent variational principle} (TDVP) on the MPS manifold~\cite{Haegeman2011,Haegeman2016}.
 
The rate at which entanglement is generated—and hence the growth of the MPS bond dimension $\chi(t)$—depends strongly on the coupling $\kappa$ and, to a lesser extent, on $\omega$. This growth limits the maximum accessible evolution time at fixed numerical accuracy, especially for large $N$. Accordingly, our simulations focus on the moderately entangled regime introduced in Sec.~\ref{section:Sec4}, where we take $\kappa=0.5$ and $\omega^2=1.5$. In this regime, the pre-quench ground state remains close to Gaussian, so the initial state is well approximated by the corresponding harmonic chain; the dominant contribution to the entanglement growth still comes from the zero mode, but compactness already affects the dynamics at comparatively early times. The qualitative lessons for the large-$\kappa$ regime should carry over unchanged, though that regime is computationally more demanding to access.

\begin{figure}[htbp]
\centering
\includegraphics[width=.68\linewidth]{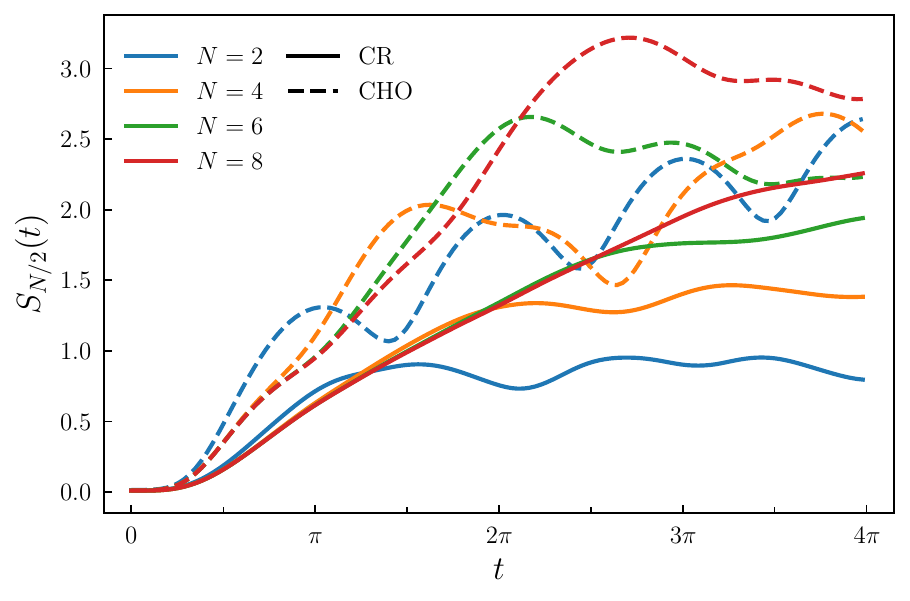}
\caption{Entanglement dynamics in the \emph{coupled-rotor} (CR) chain.
Half-chain entanglement entropy $S^{\rm CR}_{N/2}(t)$ for several system sizes $N$ as a function of $t$. At early times the entropy grows sub-extensively in $N$, while at later times compactness curbs the growth and induces saturation at a finite value.
For comparison, the \emph{corresponding harmonic chain} (CHO, dashed lines) exhibits unbounded entanglement growth.
Parameters: $\omega^2=1.5$, $\kappa=0.5$. The MPS simulations were performed with an angular momentum cutoff $M = 6$, a truncation error of $\epsilon = 10^{-6}$ and a maximal bond dimension of $\chi = 1024$.}
    \label{fig:Fig5}
\end{figure}

As shown in Fig.~\ref{fig:Fig5}, the half-chain entanglement entropy $S^{\rm CR}_{N/2}(t)$ initially grows in time $t$ with a sub-extensive dependence on system size, and therefore more slowly than in the harmonic chain. For the system sizes accessible to our simulations, both the duration of this growth regime and the subsequent saturation value $S^{\rm CR}_{N/2,\max}(N)$ increase with $N$ sub-extensively. The finite maximum is set by the compact zero mode, whose bounded configuration space and discrete spectrum prevent unbounded entanglement growth, as already seen in the two-site rotor model.

At later times, an increasing number of post-quench eigenstates with generally incommensurate frequencies contribute to the dynamics. As the many-body level spacings shrink with system size, the associated revival times are pushed to progressively longer scales and quickly move beyond the accessible simulation window, leading to a broad saturation plateau.

In this regime, one may ask whether a quasiparticle light-cone picture modified by the presence of the compact zero mode re-emerges for the rotor chain—a description that breaks down for the corresponding harmonic chain with Neumann boundary conditions under this quench, where the non-compact zero mode dominates the late-time dynamics~\cite{aimet2024experimentally}. In such a picture, entangled excitations generated by the quench would propagate with an effective maximal velocity reduced relative to the harmonic case, reflecting both the curvature of the cosine potential and the boundedness imposed by compactness. Whether this interpretation becomes sharply applicable remains an open question, however, as the limited system sizes accessible here do not allow us to establish a clear scaling regime or a well-defined light cone.

Accessing substantially larger system sizes and longer evolution times to test convergence towards a generalized Gibbs ensemble (GGE) is numerically challenging. Nevertheless, as $N$ approaches the thermodynamic limit, local observables are expected to equilibrate on average~\cite{gogolin2016equilibration}, in which case a GGE constructed from the conserved mode energies should provide an increasingly accurate description of the late-time state. In practice, however, validating this expectation directly is not currently feasible: While the GGE itself can be constructed for few-body rotor chains, computing the von Neumann entropy of its reduced density matrix rapidly becomes intractable.

\section{Implications for compact quantum field theories in ultra-cold-atom experiments}\label{section:Sec6}
Having established how compactness curbs entanglement growth in discrete rotor models, we now turn to the continuum setting relevant for ultra-cold-atom experiments, where the same physics carries over to compact quantum field theories. Starting from a (nearly) Gaussian initial state, the early-time dynamics are well captured by a non-compact Gaussian Klein--Gordon description, while at longer times the compact \emph{Tomonaga--Luttinger liquid} (TLL) governs the evolution. In particular, the compact zero mode plays the same role as in the rotor chain and ultimately caps entanglement growth.

While this picture is theoretically clear, it remains an open question whether the dynamical effects of compactness can be accessed experimentally. Existing reconstruction schemes, such as tomographical methods based on covariance matrices, assume Gaussian phase statistics and thus effectively treat the phase field—including its zero mode—as non-compact. Consequently, they are limited to early-time dynamics and cannot probe the regime where compactness becomes dynamically relevant.

\subsection{Experimental setting and low-energy field theory}
The concrete realization we have in mind is a pair of tunnel-coupled one-dimensional Bose gases of $^{87}$Rb~\cite{schweigler2017experimental, schweigler2021decay, bastianello2024sine, aimet2024experimentally}. At low energies the relative degrees of freedom are described by a compact phase field $\varphi(z)$, defined modulo $2\pi$, together with its conjugate density fluctuation $\delta\rho(z)$ on the interval $z\in[0,L]$. In the present experiments the relevant boundary conditions are approximately Neumann, $\partial_z \varphi|_{z=0,L}=0$, reflecting the box-like geometry of the system. The platform, however, is not restricted to this case. By suitable shaping of the confining and reference potentials~\cite{Tajik2019Arbitrary1DPotentials} one can also realize phase-pinned boundaries of Dirichlet type, $\varphi \approx 0 \mod 2\pi$, as well as ring geometries with periodic boundary conditions. Since the boundary conditions determine the structure of the zero-mode sector and the role of winding configurations, they form an essential ingredient of the compact-field physics discussed below.

It is important to note that in the experiment we observe the relative degrees of freedom $\frac{1}{\sqrt{2}} (\ket{\Psi_1}  - \ket{\Psi_2})$ of the two condensates $\Psi_1$ and $\Psi_2$, which is probed by matter-wave interferometry giving the relative field $\varphi(z) = \phi_1(z)-\phi_2(z)$. These have to be distinguished from the common degrees of freedom $\frac{1}{\sqrt{2}} (\ket{\Psi_1} + \ket{\Psi_2})$, which is governed by the global phase $\phi_1(z)+\phi_2(z)$. This symmetric sector is not directly visible in the interference signal, although it can in principle be reconstructed indirectly from density dynamics or with the aid of an external phase reference ~\cite{Murtadho2025a,Murtadho2025b}. The zero mode discussed below is therefore the spatially uniform mode of the relative field $\varphi(z)$, not the global common phase of the two condensates. 

In contrast to the previous sections, the experimentally relevant initial state is not simply the ground state of the pre-quench Hamiltonian. Instead, it is a low-energy state prepared by the experimental protocol, whose fluctuations reflect both the finite-temperature state of the parent condensate and additional excitations generated during the splitting or quench process itself. For a sudden splitting this includes beam-splitter (partition) noise in the relative sector~\cite{Gring2012}, whereas slower ramps allow these excitations to be shaped and partially suppressed by appropriate control strategies~\cite{Grond2009,Zhang2024,Kuriatnikov2025}.

The initial state is described by the density matrix
\begin{equation}
\rho(0) = \frac{e^{-H^{\rm sG}/(k_{\rm B} T)}}{{\rm Tr}\!\left[e^{-H^{\rm sG}/(k_{\rm B} T)}\right]}\,,   
\end{equation}
with $T$ the temperature and $k_{\rm B}$ the Boltzmann constant. The sine–Gordon Hamiltonian $H^{\rm sG}$ reads
\begin{equation}
H^{\rm sG} = \int_{0}^{L} dz \left[
        g_{\mathrm{1D}}\, \delta \rho(z)^2 
        + 2 \hbar J n_{\mathrm{1D}}\,\bigl(1 - \cos \varphi(z)\bigr)
        + \frac{\hbar^2 n_{\mathrm{1D}}}{4m}\, \bigl(\partial_z \varphi(z)\bigr)^2
    \right],
\end{equation}
where $n_{\rm 1D}$ is the linear density, $m$ the atomic mass, $J$ the tunnel coupling, and $g_{\rm 1D}$ the one-dimensional interaction strength. As in the protocol of Ref.~\cite{schweigler2021decay}, the tunnel coupling is quenched to zero at time $t=0$, $J\to0$. The subsequent dynamics are therefore governed by the post-quench Hamiltonian
\begin{equation}
H^{\rm TLL} = \int_{0}^{L} dz \left[
        g_{\mathrm{1D}}\, \delta \rho(z)^2
        + \frac{\hbar^2 n_{\mathrm{1D}}}{4m}\, \bigl(\partial_z \varphi(z)\bigr)^2
    \right],
\end{equation}
which is precisely the compact TLL Hamiltonian with identification $\varphi\equiv\varphi+2\pi$.

This quench protocol has been realized in ultra-cold-atom experiments, with Ref.~\cite{aimet2024experimentally} in particular measuring the time evolution of the von Neumann entropy of a subsystem after the quench.
\medskip
As in the rotor chains, one could in principle consider initial states deep in the non-Gaussian regime of the sine–Gordon Hamiltonian, where compactness is already essential at $t=0$. Here, however, we focus on the opposite regime directly relevant for the experiments: A deep quench from a state that is extremely well approximated by the thermal state of a massive Klein–Gordon Hamiltonian,
\begin{equation}
H^{\rm KG}
=
\int_{0}^{L} \! dz\,
\left[
g_{\rm 1D}\,\delta\rho(z)^2
+
\hbar J n_{\rm 1D}\,\varphi(z)^2
+
\frac{\hbar^2 n_{\rm 1D}}{4m}\,
\bigl(\partial_z \varphi(z)\bigr)^2
\right],
\end{equation}
in which the field is treated as non-compact, $\varphi(z)\in\mathbb{R}$.
Strictly speaking, the experimental system always realizes a compact phase field originating from the underlying sine–Gordon description. The harmonic Klein–Gordon approximation is valid in the regime where the phase fluctuations remain small, such that $\langle\cos\varphi\rangle\approx1$ and the cosine potential is explored only near its minimum. While the compact nature of the field still permits $2\pi$ phase slips associated with topological excitations, these processes are strongly suppressed in the regime considered here and do not affect the Gaussian structure of the initial state. Consequently, the initial correlations and entanglement properties are indistinguishable from those of the non-compact Klein–Gordon theory.

\subsection{Mode decomposition and the compact zero mode}
As in the discrete models, both the initial massive Klein--Gordon Hamiltonian and the post-quench Hamiltonians—compact or non-compact—decouple mode by mode when expressed in momentum space. For Neumann boundary conditions it is natural to expand the fields as
\begin{equation}
\varphi(z)=\sum_{k=0}^{\infty}f_k(z) \varphi_k,\qquad \delta\rho(z)=\sum_{k=0}^{\infty}f_k(z)\delta\rho_k
\end{equation}
with basis functions
\begin{equation}
    f_k(z)=
    \begin{cases}
        \displaystyle \frac{1}{\sqrt{L}}, & k=0,\\[6pt]
        \displaystyle \sqrt{\frac{2}{L}}\cos\!\left(\frac{k\pi z}{L}\right), & k\ge1,
    \end{cases}
\end{equation}
which satisfy the orthonormality relation
\(\int_0^{L}\!dz\,f_k(z)f_l(z)=\delta_{k,l}\). The mode operators follow as
\begin{equation}
\varphi_k=\int_0^{L} dz\,f_k(z)\varphi(z),\qquad
\delta\rho_k=\int_0^{L} dz\,f_k(z)\delta\rho(z),
\end{equation}
with $[\varphi_k,\delta\rho_l]=i\,\delta_{k,l}$. Rescaling the Hamiltonian as $\tilde H=H/(2g_{\rm 1D})$ correspondingly rescales time as $\tilde t = 2g_{\rm 1D}\, t$, the Hamiltonian reduces to a sum of decoupled harmonic oscillators. For the rescaled pre-quench Hamiltonian one finds
\begin{equation}
\tilde H^{\rm KG}
=\frac12\sum_{k=0}^{\infty}
\left[
\delta\rho_k^2+\Omega_{i,k}^2\,\varphi_k^2
\right],
\qquad
\Omega_{i,k}^2
=\frac{\hbar J n_{\rm 1D}}{g_{\rm 1D}}
+\frac{\hbar^2 n_{\rm 1D}}{4m g_{\rm 1D}}
\left(\frac{\pi k}{L}\right)^2 .
\end{equation}
After the quench $J\to0$, the Hamiltonian becomes
\begin{equation}
\tilde H^{\rm TLL}
=\frac12\sum_{k=0}^{\infty}
\left[
\delta\rho_k^2+\Omega_{f,k}^2\,\varphi_k^2
\right],
\qquad
\Omega_{f,k}^2
=\frac{\hbar^2 n_{\rm 1D}}{4m g_{\rm 1D}}
\left(\frac{\pi k}{L}\right)^2 .
\end{equation}
In contrast to the coupled-rotor system of Sec.~\ref{section:Sec3} and ~\ref{section:Sec5}, the continuum field theory factorizes exactly into independent mode sectors, with no global parity constraints.

For the non-compact Klein--Gordon theory, all mode coordinates $\varphi_k$ take values in $\mathbb{R}$. By contrast, in the compact theory with $\varphi\equiv\varphi+2\pi$, only the spatially uniform mode $k=0$,
\begin{equation}
\varphi_0=\frac{1}{\sqrt{L}}\int_0^L dz\,\varphi(z),
\end{equation}
inherits the compact topology of the original field. It satisfies the identification
\begin{equation}
\varphi_0\equiv \varphi_0+2\pi R_0,
\qquad
R_0:=\sqrt{L},
\end{equation}
and thus corresponds to a quantum rotor with compactification radius $R_0=\sqrt{L}$, while all higher modes $\varphi_{k\ge1}$ remain non-compact harmonic oscillators.

After the quench, this spatially uniform mode ($k=0$) becomes a zero mode. In the compact theory its angular nature implies a discrete momentum spectrum, $\delta\rho_0=p/R_0$ with $p\in\mathbb{Z}$, whereas in the non-compact theory the zero-momentum mode lives on $\mathbb{R}$ and has a continuous spectrum.

\subsection{Estimating compactness timescale}
\label{subsection:Sec6_3}
We consider the quench dynamics probed in current experiments, where a Gaussian thermal initial state of a massive Klein--Gordon theory is evolved under a massless post-quench Hamiltonian. For experimentally accessible evolution times, the dynamics are well described by a massless, non-compact Klein--Gordon theory obtained from $H^{\rm KG}$ by setting $J=0$. Its Hamiltonian density coincides with that of the Tomonaga--Luttinger liquid, except that the field $\varphi$ takes values in $\mathbb{R}$ rather than on a circle.

At early times, compactness is effectively irrelevant: both the initial state and subsequent dynamics lie in a regime where the sinusoidal potential is indistinguishable from its quadratic approximation. The non-compact theory therefore provides an accurate effective description. However, since the field is fundamentally compact, $\varphi \equiv \varphi + 2\pi$, this description must eventually break down. At sufficiently long times, the spreading of the global (zero) mode causes the phase distribution to explore the compact manifold, rendering compactness dynamically relevant. We now estimate the timescale for this crossover.

We consider the experimental parameters of Ref.~\cite{aimet2024experimentally},
\begin{equation}
\begin{aligned}
L &= 49\,\mu{\rm m},\qquad
n_{\rm 1D} = 70\,\mu{\rm m}^{-1},\qquad
g_{\rm 1D} = 8.594\times 10^{-39}\,{\rm kg\,m^3 s^{-2}},\\
m &= 1.433\times 10^{-25}\,{\rm kg},\qquad
J = 2\pi\times 0.76\,{\rm Hz},\qquad
T = 49\,{\rm nK}.
\end{aligned}
\label{eq:exp_parameters}
\end{equation}
The following estimate is intended only at the level of an order-of-magnitude, back-of-the-envelope calculation. The experiment involves inhomogeneities and other effects not captured by the homogeneous model, and the onset of compactness is inferred from dynamics derived within a non-compact theory. The resulting timescale should therefore be viewed as a rough guide; nevertheless, it suffices to assess whether compactness effects are within experimental reach.

Since compactness is governed by the global (zero) mode, we estimate the compactness timescale by requiring that the zero-mode variance reaches that of a uniform distribution on its compact domain of width $2\pi R_0$,
\begin{equation}
\sigma_{\varphi_0}^2(\tilde t_{\rm c})
\sim
\frac{\pi^2 R_0^2}{3}.
\label{eq:compactness_criterion_app}
\end{equation}
This criterion likely overestimates the onset time even if we assumed a compact modeling for the time evolution to obtain the estimate, since compactness effects may already become relevant once the spreading wavefunction wraps around the circle and its tails overlap. At the level of an order-of-magnitude estimate, however, this distinction is not resolved.

In the non-compact theory, following a quench to $\omega_f=0$, the zero-mode variance evolves as
\begin{equation}
\sigma_{\varphi_0}^2(\tilde{t})=\sigma_{\varphi_0}^2(0)+\left(\frac{\tilde{t}}{\hbar}\right)^2 \sigma_{\delta\rho_0}^2(0).
    \label{eq:sigmaQ0_free}
\end{equation}
For a thermal initial state of the pre-quench Klein--Gordon Hamiltonian, the initial zero-mode variances are
\begin{equation}
\sigma_{\varphi_0}^2(0)=\frac{1}{2\Omega_{i,0}}\coth\!\left(\frac{g_{\rm 1D}\Omega_{i,0}}{k_{\rm B}T}\right),
\qquad
\sigma_{\delta\rho_0}^2(0)=\frac{\Omega_{i,0}}{2}\coth\!\left(\frac{g_{\rm 1D}\Omega_{i,0}}{k_{\rm B}T}\right),
\label{eq:thermal_variances}
\end{equation}
with $\Omega_{i,0}=\sqrt{\frac{\hbar J n_{\rm 1D}}{g_{\rm 1D}}}$.

Combining Eqs.~\eqref{eq:sigmaQ0_free} and \eqref{eq:thermal_variances}, the compactness timescale $\tilde t_{\rm c}$ defined by
$\sigma_{\varphi_0}^2(\tilde t_{\rm c})\sim \pi^2 R_0^2/3$
simplifies in the deep-quench regime,
$\sigma_{\varphi_0}^2(0)\ll \pi^2 R_0^2/3$, to
\begin{equation}
\tilde t_{\rm c}
\sim
\frac{\hbar}{\sigma_{\delta\rho_0}(0)}\frac{\pi R_0}{\sqrt{3}}
=
\frac{\hbar}{
\sqrt{\frac{\Omega_{i,0}}{2}
\coth\!\left(\frac{g_{\rm 1D}\Omega_{i,0}}{k_{\rm B}T}\right)}
}\frac{\pi R_0}{\sqrt{3}}\,.
\label{eq:tc_simple}
\end{equation}

Finally, expressing the result in terms of physical time
$t=\tilde t/(2g_{\rm 1D})$
and substituting $R_0=\sqrt{L}$ and
$\Omega_{i,0}=\sqrt{\hbar J n_{\rm 1D}/g_{\rm 1D}}$,
we obtain
\begin{equation}
t_{\rm c}
\sim
\frac{\hbar\pi}{g_{\rm 1D}\sqrt{6}}
\sqrt{L}
\left(\frac{g_{\rm 1D}}{\hbar J n_{\rm 1D}}\right)^{1/4}
\sqrt{\tanh\!\left(
\frac{\sqrt{g_{\rm 1D} \hbar J n_{\rm 1D}}}{k_{\rm B}T}
\right)}.
\label{eq:tc_final}
\end{equation}
For the parameters in Eq.~\eqref{eq:exp_parameters}, this yields a compactness timescale of $t_{\rm c}\approx 12\,{\rm ms}$, which is comparable to experimentally accessible evolution times.

This estimate should be interpreted with some care. Apart from what has been said already, it isolates the zero mode, whereas the experimental dynamics involve thermally populated higher modes. These modes may modify the redistribution of phase fluctuations and thus shift the precise onset of compactness effects.

In the main part of this work we focused on a zero-mode–dominated regime, which provides the cleanest setting and allows for analytical results. Whether such a regime can be cleanly isolated in experiments remains an open question. For nonzero modes $k\ge1$, the relative importance of mass and gradient terms is characterized by
\begin{equation}
r_k=\sqrt{1+\frac{4m J}{\hbar (\frac{\pi k}{L})^2}}\,.
\end{equation}
Freezing out the relative modes would require \(r_k\simeq1\), corresponding to
\(J\ll \hbar\pi^2/(4mL^2)\), which is not satisfied for the experimental parameters considered here. The post-quench evolution therefore involves both the zero mode and higher momentum modes. Nevertheless, in the non-compact description the zero mode still dominates the entanglement growth, since its variance grows unboundedly in time whereas the higher modes produce only bounded oscillatory contributions.

Ultimately, a quantitative determination of the onset of compactness effects—including their impact on observables such as the entanglement ceiling $S_{L/2,\rm max}$, as well as regimes where the initial state lies deep in the compact sine–Gordon phase—requires a treatment within the fully compact field theory. As in the rotor chain, one expects long-time equilibration towards a generalized Gibbs ensemble (GGE), but determining the corresponding stationary entanglement entropy requires numerical simulations of the compact theory and lies beyond the scope of the present work.

\subsection{Experimental observation of zero-mode spreading and associated challenges}
For the quench considered in this work—from a phase-locked initial state, well described by a thermal state of a massive Klein–Gordon Hamiltonian, to Tomonaga–Luttinger-liquid dynamics—the compact nature of the phase field, defined modulo $2\pi$ implies that the spreading of the zero mode, and hence the associated entanglement growth, ultimately saturates. Independently of this theoretical prediction, however, the compactness of the phase field also introduces practical challenges for its experimental reconstruction.

In matter-wave interferometry experiments the relative phase profile $\varphi(z)$ is reconstructed locally from the positions of interference fringes and is therefore reported in the principal interval $[-\pi,\pi)$. When analyzing such data one may either (i) \emph{wrap} the phase, keeping it pointwise within the principal interval $[-\pi,\pi)$, or (ii) \emph{unwrap} it by lifting the profile to the universal cover $\mathbb{R}$ using a spatial continuity criterion along $z$ within a single experimental realization.

Phase unwrapping is not always possible, as it requires the phase difference between neighboring samples to remain sufficiently small~\cite{itoh1982analysis}. Within a single experimental shot, when the \textit{spatial} resolution is adequate, the phase profile can be followed continuously along the system and unwrapping becomes physically well defined. In this regime, the reconstructed unwrapped phase profile may extend well beyond the interval $[-\pi,\pi)$, preserving the integer-valued winding content of the phase field and making localized $2\pi$ phase slips explicit, corresponding to sine--Gordon solitons as demonstrated in Ref.~\cite{schweigler2017experimental}. By contrast, enforcing pointwise wrapping of the phase can obscure such topological features by introducing artificial discontinuities at the boundaries of the principal interval.

If the spatial resolution is insufficient, however, unwrapping becomes unreliable. When the resolution length exceeds the characteristic size of topological excitations such as solitons, or even the coherence length of the system, the phase difference between neighboring samples may violate the unwrapping criterion and the unwrapped phase cannot be uniquely determined.

A second limitation arises in the opposite regime when the spatial resolution becomes so high that the number of atoms contained within a resolution element becomes very small, of order unity. In this case the uncertainty in estimating the local phase becomes comparable to the phase itself, and the effective emergent description underlying the unwrapping procedure breaks down, requiring a more microscopic description of the system. In practice the transition between regimes where the phase can be reliably followed and unwrapped and those where this becomes impossible is not sharp, but rather a broad crossover region in which phase unwrapping becomes unreliable and may produce random $\pm\pi$ errors.

While compact features of the phase field can thus be accessed within a single experimental realization, tracking the dynamical effects of compactness in time is considerably more challenging. The reason is that measurements in cold-atom interferometry are destructive: phase profiles recorded at different evolution times correspond to independent realizations rather than successive snapshots of a single trajectory.

The contrast becomes clear by comparison with a classical system. Consider, for example, a chain of coupled pendula whose phases are initially locked by a strong gravitational field and then evolve freely once the field is switched off. If the motion were recorded with sufficiently fine \textit{temporal} resolution, the full phase evolution could be reconstructed despite the $2\pi$ ambiguity of each instantaneous measurement since temporal continuity fixes the branch of the compact phase and allows the dynamics to be lifted to a continuous trajectory on the universal cover. In cold-atom interferometry experiments such procedure is not available. Although spatial continuity allows the profile within a single realization to be unwrapped, temporal continuity is (currently) not available in these experiments and cannot be used to reconstruct a globally unwrapped phase evolution.

This distinction between wrapped and unwrapped descriptions is therefore essential when interpreting experimental phase distributions. While spatial unwrapping can reveal the underlying phase structure within a single realization, the experimentally accessible distributions ultimately reflect the compact phase variable, and care must be taken in relating them to the underlying dynamics on the universal cover.

To make this issue concrete we focus on the zero mode of the phase field, defined as the spatial average
$\varphi_0(t)=\frac{1}{L}\int_0^L dz\,\varphi(z,t)$. We treat $\varphi_0$ as a dimensionless phase variable. In contrast to the normalization used in previous subsections, its compactification radius is therefore $R_0=1$, so that $\varphi_0\in[-\pi,\pi)$ modulo $2\pi$.

\begin{figure}[htbp]
\centering
\includegraphics[width=1.\linewidth]{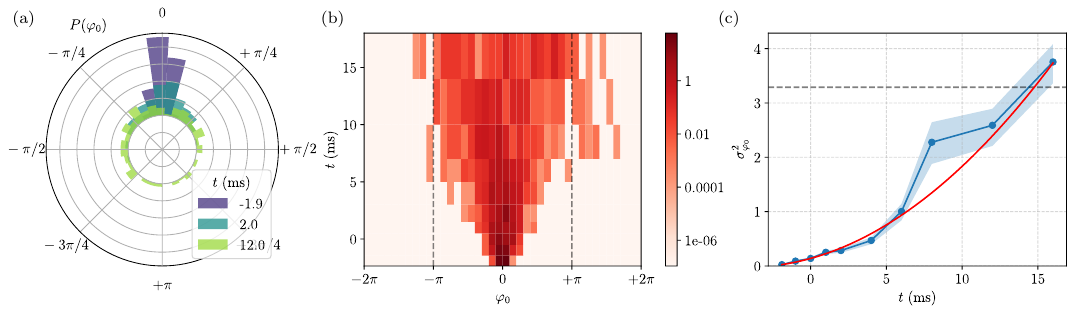}
\caption{Experimental observation of spreading of the zero-mode distribution after a quantum quench. Polar (a) and density plot (b) histograms of the zero-mode phase values at different times (before, soon after and long after the ramp). Starting from a zero-mode phase-locked state, the TLL dynamics cause spreading of the distribution.
Once the distribution spans the entire circle, inferring information about further spreading of the distribution from the experimental measurements is challenging. 
(c) Time evolution of the zero-mode variance and parabolic fit of the data up to the time when the variance of a uniform distribution (dashed horizontal line) is reached. 
}
    \label{fig:Fig6}
\end{figure}

We now consider measurements of the dynamical spreading of this zero mode following the quench. For a set of parameters similar to those discussed above (though not identical), the experimentally reconstructed zero-mode phase distribution $P(\varphi_0)$ is shown in Fig.~\ref{fig:Fig6}. Panels (a) and (b) display polar and density histograms at representative evolution times, while panel (c) shows the time evolution of the variance $\sigma_{\varphi_0}^2(t)$.

Starting from a sharply localized distribution, the zero mode spreads rapidly after the quench. At early times the phase distribution remains confined to a narrow angular sector and the variance grows approximately quadratically in time, consistent with the ballistic spreading predicted by the non-compact massless theory [cf.\ Eq.~\eqref{eq:sigmaQ0_free}]. In this regime, the phase profile remains sufficiently localized that it can be reliably unwrapped and the associated winding numbers extracted.

As the evolution proceeds the distribution spreads across an increasingly large fraction of the interval $[-\pi,\pi)$. The density histograms in Fig.~\ref{fig:Fig6}(b) already show the edges of the principal interval beginning to populate after about $t\approx 5\,{\rm ms}$. Once the distribution spans the full interval the phase can no longer be reliably unwrapped. The underlying winding information becomes inaccessible and further spreading on the universal cover is effectively folded back into $[-\pi,\pi)$. The dynamics can therefore only be analyzed in terms of the wrapped phase. 

Notably, the variance reaches the value $\pi^2/3$ corresponding to a uniform circular distribution already at $t\approx15\,{\rm ms}$ in Fig.~\ref{fig:Fig6}(c), so the compactness timescale $t_c$ estimated in Sec.~\ref{subsection:Sec6_3} seems to be reasonably accurate. Panel (a) shows that the distribution at this time is still far from uniform. This is consistent with the fact that the estimate of Sec.~\ref{subsection:Sec6_3} was obtained from the non-compact zero-mode dynamics and therefore assumes continued spreading on $\mathbb{R}$. As emphasized there, the estimate should be interpreted as a conservative upper bound, and the discrepancy observed here can itself be viewed as a signature of the compact nature of the phase and the onset of compactness effects.

Reconstructing the continued spreading beyond this point is experimentally nontrivial because the phase is measured only modulo $2\pi$. The compact zero mode $\varphi_0$ is an angular variable on $S^1$, while a corresponding ``running'' variable $\varphi_0^{\rm run}\in\mathbb{R}$ can be defined only together with an integer winding record $W$ that counts crossings of the branch cut, $\varphi_0^{\rm run}(t)=\varphi_0(t)+2\pi W(t)$.

Access to this winding record would be required to reconstruct the evolution on the universal cover. In classical dynamics this information could in principle be obtained from sufficiently frequent, non-destructive snapshots that allow the winding number $W(t)$ to be tracked. In the quantum experiments considered here, however, each time point corresponds to an independent realization.

Once the phase distribution wraps around the circle, the winding information cannot easily be recovered across shots and only the compact marginal $P(\varphi_0\bmod2\pi)$ remains operationally well defined. Consequently the spreading of the phase distribution after the quench can be quantitatively tracked only at early times, while the distribution remains localized. At later times wrapping sets in, and the subsequent compact dynamics cannot be directly resolved within the present measurement protocol.

Whether the information of the compact theory is experimentally accessible once the phase distribution has wrapped around is left for future investigation, possibly enabling more refined tomographical reconstruction schemes that allow probing late-time regimes reliably where compactness plays a dynamical role.

The compact description thus becomes operationally unavoidable in two distinct situations: spatially, when the resolution is insufficient to follow the phase continuously and reconstruct winding numbers reliably within a single shot; and dynamically, when the measurements are destructive and temporal continuity cannot be used to track the winding across independent realizations. In both cases the phase is only accessible as a compact variable on $S^1$, and the non-compact description ceases to be operationally meaningful. These are precisely the conditions under which the late-time dynamics — including the saturation of entanglement growth predicted by the compact theory — should become manifest, provided reconstruction schemes adapted to the compact phase are employed.

As a final remark, while the experiments discussed above effectively realize Neumann boundary conditions, engineered pinning of the relative phase at the edges (Dirichlet-type conditions) provides a complementary probe of compactness. In such settings the total phase drop across the system is restricted to discrete $2\pi$ winding sectors, and unwrapping the spatial phase profile within a single experimental realization directly reveals the corresponding integer winding number that would otherwise be obscured by pointwise wrapping.

\subsection{Lattice discretization and connection to rotor chains}
In the parameter regime discussed in the previous subsection, the initial state is a thermal state rather than the ground state of the pre-quench Hamiltonian. Nevertheless, as a first step toward theoretical insight, it is useful to approximate the initial state by the ground state and thus describe the subsequent dynamics at the level of a pure-state wave function. This simplification captures the essential structure of the quench dynamics, while a more complete treatment would retain the full thermal character of the initial state.

Within this approximation, the time-evolved state is fully characterized by its wave function, allowing, in principle, for a direct computation of the entanglement entropy of a spatial bipartition. One possible strategy is to truncate the mode expansion at a finite cutoff \(N\) and reconstruct the fields on a discrete spatial grid \(\{z_n\}_{n=0}^{N-1}\) via
\begin{equation}
\varphi(z_n) = \sum_{k=0}^{N-1} f_{nk} \varphi_k,
\qquad
\delta\rho(z_n) = \sum_{k=0}^{N-1} f_{nk} \delta\rho_k,
\end{equation}
with discrete basis functions
\begin{equation}
f_{nk} =
\begin{cases}
\sqrt{\dfrac{1}{L}}, & k = 0,\\[2pt]
\sqrt{\dfrac{2}{L}} \cos\left[\dfrac{(n + \tfrac{1}{2}) k \pi}{N}\right], & k \ge 1,
\end{cases}
\end{equation}
and then use the resulting lattice wave function $\Psi(\{\varphi(z_n)\};t)$ to construct and diagonalize the reduced density matrix of a spatial subsystem.

An alternative, and conceptually more direct route is to discretize the position-space Hamiltonian from the outset. We replace the continuum interval $[0,L]$ by a one-dimensional lattice of $N$ sites with spacing $a=L/N$ and introduce canonical pairs $(\varphi(z_n),\delta\rho(z_n))$ at each site $n=0,\dots,N-1$.

A crucial subtlety is the discretization of the gradient term $(\partial_z\varphi)^2$. For a non-compact field one might simply take
\begin{equation}
(\partial_z\varphi(z_n))^2 \simeq \frac{(\varphi(z_{n+1})-\varphi(z_n))^2}{a^2}\,,
\end{equation}
but this breaks the $2\pi$-periodicity of $\varphi$ and is,  therefore, inappropriate for the compact theory. Instead, we adopt a discretization that preserves compactness,
\begin{equation}
(\partial_z\varphi(z_n))^2 \simeq \frac{2}{a^2}\bigl[1-\cos(\varphi(z_{n+1})-\varphi(z_{n}))\bigr],
\end{equation}
which reduces to the harmonic form only in the small-angle limit.

With this choice, the lattice sine–Gordon Hamiltonian becomes
\begin{equation}
    H^{\rm sG}=\lim_{a\to 0}\,\frac{a}{2}\left[ g_{\mathrm{1D}}\sum_{n=0}^{N-1}
\delta\rho(z_{n})^2+2 \hbar J n_{\mathrm{1D}}\sum_{n=0}^{N-1}\bigl(1-\cos\varphi(z_n)\bigr)+\frac{\hbar^2 n_{\rm 1D}}{2m a^2}\sum_{n=0}^{N-2}
\bigl[1-\cos(\varphi(z_{n+1})-\varphi(z_n))\bigr]\right],
\end{equation}
while, after the quench, the TLL Hamiltonian reads
\begin{equation}
    H^{\rm TLL}=\lim_{a\to 0}\,\frac{a}{2}\left[ g_{\mathrm{1D}}\sum_{n=0}^{N-1}
\delta\rho(z_n)^2+\frac{\hbar^2 n_{\rm 1D}}{2m a^2}\sum_{n=0}^{N-2}
\bigl[1-\cos(\varphi(z_{n+1})-\varphi(z_n))\bigr]\right],
\end{equation}
Thus the coupled rotor chain of Sec.~\ref{section:Sec5} appears naturally as the compact and correctly discretized version of the continuum field theory in the limit $a\to 0$. In this lattice formulation, the periodicity of the field is manifest.

Given that, for the two-site rotor model, the compact zero mode could be isolated explicitly and $S^{\rm CR}_{1,\rm max}$ estimated analytically via the single-site GGE entropy $S^{\rm CR}_{1,{\rm GGE}}$, it is natural to ask whether analogous analytic simplifications are possible for the many-body rotor chain in the continuum limit. Addressing this question, and thereby obtaining analytic control over $S_{L/2,\rm max}$ and its scaling in the compact TLL, could be an interesting starting point for future investigations.
We further note that the compact boson on a finite interval admits several alternative discretizations, in either coordinate or momentum space, that provide options for numerical treatment (see, e.g., Ref.~\cite{roy2021quantum} and Refs.~\cite{Horvath2022, murciano2022postquantum, schmoll2023hamiltonian} for a lattice discretization and momentum space truncation, respectively), and its entanglement and quench dynamics have also been analyzed directly in boundary conformal field theory~\cite{Bastianello2019, estienne2024entanglement}.

\section{Conclusions and outlook}\label{section:Sec7}
In this work we have investigated the dynamical role of compactness in shaping
late-time entanglement growth in bosonic systems. Our analysis focuses on the
zero mode that emerges after a global quench to vanishing on-site potential
frequency. The central finding is a conceptual distinction that has largely
been overlooked in the literature on entanglement dynamics: It is not the
presence of a zero mode \emph{per se} that drives logarithmic entanglement
growth, but specifically its \emph{non-compact} character.

A zero mode defined on an unbounded configuration space behaves as a free
particle. Its wavefunction spreads without bound, and its continuous spectrum
leads to indefinite dephasing, resulting in logarithmically diverging
entanglement growth. By contrast, when the zero mode is compact, the dynamics
are qualitatively different. The bounded configuration space restricts spatial
spreading and the discrete spectrum prevents indefinite dephasing, causing the
entanglement entropy to saturate at a finite value.

More generally, the mechanism responsible for this saturation does not rely on
the presence of a zero mode. Entanglement growth is ultimately bounded whenever
the relevant degrees of freedom evolve within a confined configuration space.
Such confinement may arise either from compactness or from an explicit
confining potential, whichever mechanism dominates. While both mechanisms lead to saturation, compactness
imposes a stricter constraint by limiting the accessible configuration space
itself, however potential confinement may come into play before these limits are reached. In the context of the sine-Gordon quench problem, compactness is therefore expected to be important whenever the potential is too shallow to confine the spreading of the center-of-mass. The zero-mode quench studied here provides a particularly transparent though not the only
realization of this mechanism: In the absence of any confining potential,
saturation arises solely from compactness.

We demonstrated this mechanism at increasing levels of complexity by comparing
compact and non-compact realizations side by side. Throughout our analysis we focus on a zero-mode–dominated regime, where the early-time dynamics are well captured by the Gaussian non-compact approximation.

In the minimal two-site setting, a direct comparison of coupled harmonic
oscillators and coupled quantum rotors exposes the physics in its cleanest
form. While the early-time dynamics of the two systems are practically
indistinguishable, their long-time behavior differs qualitatively. The
non-compact zero mode of the harmonic oscillator spreads without bound and
produces logarithmically growing entanglement, whereas the compact rotor
degree of freedom wraps around its bounded domain and leads to saturation of
the entropy, whose ceiling $S_{1,\mathrm{max}}^{\mathrm{CR}}$ we estimated
analytically via the generalized Gibbs ensemble. Extending the analysis to
$N$-site chains using matrix product state simulations confirms that the same
mechanism governs the many-body setting: The entanglement entropy grows
sub-extensively and ultimately saturates at a finite value whose height and
onset time increase with system size, in stark contrast to the harmonic chain
where unbounded growth persists.

Connecting these insights to experiment, we showed that the coupled-rotor chain emerges naturally as the compact lattice discretization of the Tomonaga–Luttinger liquid, and that an experimentally realized quench from a
massive Klein–Gordon initial state to the compact TLL reproduces precisely the
same compact zero-mode physics at late times.

For the parameter regime of current experiments, the compactness timescale is analytically
estimated to be on the order of $10\,\mathrm{ms}$, placing the compact regime within presently accessible evolution times. However, this timescale is not
fixed: Reducing the system length $L$, or tuning the tunnel coupling $J$ or the linear density
$n_{1\mathrm{D}}$ provides experimental handles to shorten the onset of compactness effects further.
Moreover, the estimate should be interpreted with care. It is obtained by extrapolating the non-compact zero-mode dynamics, whereas the true dynamical hallmark of compactness—the saturation of zero-mode spreading 
and of the associated entanglement entropy—must ultimately be inferred from the compact phase distribution $P(\varphi_0)$. In this respect, the preliminary experimental data presented in Fig.~\ref{fig:Fig6} already show clear evidence of ballistic spreading of the zero-mode phase distribution following the quench, providing a direct signature of the early-time dynamics preceding the compact regime.

Accessing the saturation of this spreading—the experimental fingerprint of
compactness—therefore constitutes a concrete near-term experimental target.
Achieving this will likely require either improved experimental parameters
along the lines described above or non-destructive imaging protocols that
exploit temporal continuity to reconstruct the winding content of the phase
field across shots. A central challenge is that standard tomographic reconstruction schemes, based on covariance matrices assume non-compact Gaussian phase statistics and therefore cannot access the regime where compactness becomes dynamically relevant. Extending such methods to account for the periodic nature of the phase field would thus be an essential prerequisite for probing the compact regime experimentally and could support ongoing efforts to detect entanglement using entanglement witnesses~\cite{CVWitness}.

Several important open problems remain. On the analytical side, the generalized
Gibbs ensemble (GGE) provides an accurate estimate of the entropy ceiling
$S_{1,\mathrm{max}}^{\mathrm{CR}}$ in the two-site setting. However, computing
the von Neumann entropy of the GGE reduced density matrix rapidly becomes
intractable as the system size increases. Obtaining analytical control over
the entanglement ceiling $S_{L/2,\mathrm{max}}$ and its scaling with system
size in the compact Tomonaga–Luttinger liquid—ideally connecting to results
from boundary conformal field theory~\cite{Bastianello2019,estienne2024entanglement}—
is therefore a natural and important next step.

On the numerical side, accessing larger system sizes and longer evolution
times in the rotor-chain simulations remains computationally demanding.
More generally, developing numerical methods capable of treating compact
field-theory quenches beyond the Gaussian approximation—capturing the
interplay between harmonic fluctuations and genuinely compact degrees of
freedom—remains an important challenge. In particular, simulating quenches
from initial states deep in the non-Gaussian sine–Gordon phase, where
compactness already plays an essential role at $t=0$, and understanding how
the entanglement ceiling $S_{L/2,\mathrm{max}}$—expected to coincide with the
corresponding GGE value—scales in such non-perturbative regimes remain
computationally demanding but physically important targets.

More broadly, our findings highlight how compactness and topology can
qualitatively constrain late-time dynamics far from equilibrium. The mechanism
identified here—compact configuration space acting as a structural bound on
entanglement growth—is not specific to cold-atom platforms. Compact bosonic
degrees of freedom arise naturally in lattice gauge theories, models of
quantum gravity, and a variety of condensed-matter systems with angular or
topological variables. Investigating whether analogous compactness-induced saturation of entanglement occurs in these settings, and whether it leaves observable signatures, may reveal a broader class of universal constraints on entanglement growth far from equilibrium.

\section*{Acknowledgements}
We thank Leo Shaposhnik and Shozab Qasim for helpful discussions.

\paragraph{Funding information}
This work has been supported by the DFG Research Unit FOR 2724 on ‘Thermal machines in the quantum world’, Berlin Quantum, 
the Munich Quantum Valley, the Quantum Flagship (PasQuans2, Millenion),
the BMFTR (MuniQC-Atoms), and the European Research Council (DebuQC). J.\,S.\, acknowledges support by the CoE: QuantA and by the ERC-AdG: \textit{Emergence in Quantum Physics} (EmQ).

\begin{appendix}
\numberwithin{equation}{section}

\section{Bounding and estimating entanglement in the two-site rotor model}
\subsection{Uniform-in-time upper bound on $S^{\rm CR}_1$}\label{app:bound}
Here we derive a rigorous, uniform-in-time upper bound on the entanglement entropy $S_1^{\rm CR}(\cdot)$ generated after the quench in the two-site rotor model. The derivation is fully general and applies equally to the rotor chain and to its continuum limit—the compact Tomonaga–Luttinger liquid. The key observation is that if the post-quench Hamiltonian is bounded below by a positive operator acting solely on subsystem~1, then the reduced density matrix $\rho_1(.)$ has a uniformly bounded mean energy with respect to this operator for all times. By the Gibbs variational principle—which states that the Gibbs state maximizes entropy under a fixed energy constraint—this immediately implies a finite upper bound on $S_1(.)$ that is uniform in time.

\begin{figure}[htpb]
    \centering
    \includegraphics[width=.9\linewidth]{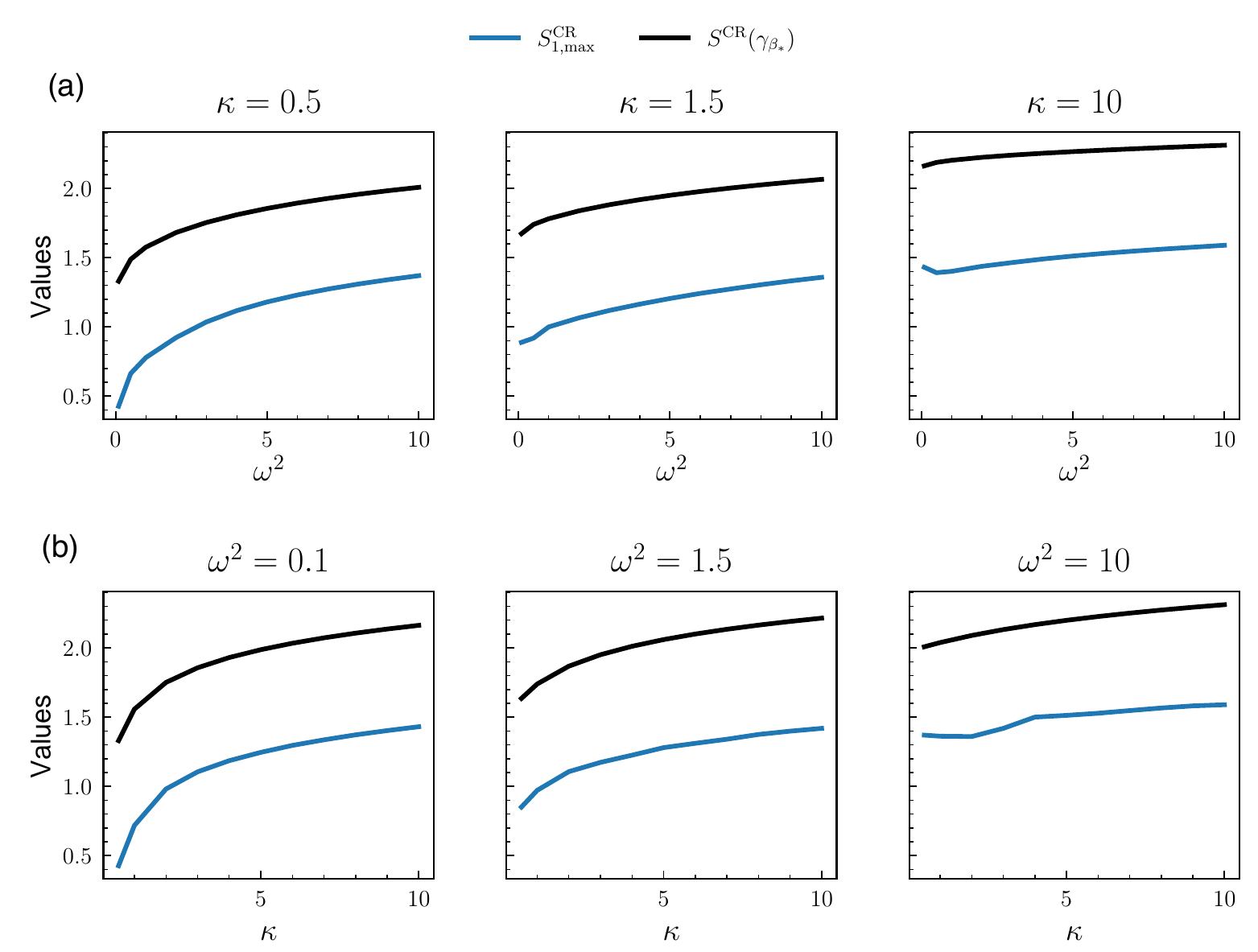}
    \caption{Maximal bipartite entanglement entropy in the two-site \emph{coupled rotor} (CR) model and a uniform-in-time upper bound.
Shown is the numerically obtained maximum $S^{\rm CR}_{1,\max}$ together with the bound derived from the Gibbs variational principle.
(a) Dependence on the initial trap strength $\omega^2$ at fixed $\kappa$.
(b) Dependence on the coupling $\kappa$ at fixed $\omega^2$.}
    \label{fig:Fig7}
\end{figure}

\medskip
Let $\mathcal H=\mathcal H_1\otimes\mathcal H_2$ be a bipartite Hilbert space and $\ket{\psi(t)}=e^{-i H_f t}\ket{\psi(0)}$ the post-quench state vector. Assume that the Hamiltonian satisfies the operator inequality
\begin{equation}
H_f \ge h_1\otimes \mathbb I_2 ,
\label{eq:local-lower-bound-derivation}
\end{equation}
for some positive operator $h_1$ acting only on subsystem~1.  
In the two-site rotor model this condition is realized by choosing $h_1=\tfrac12 p_1^2$, since the interaction term $1-\cos(x_1-x_2)$ is non-negative.

Let $E_{\mathrm{tot}}=\langle\psi(0)|H_f|\psi(0)\rangle$ denote the conserved total energy. From Eq.~\eqref{eq:local-lower-bound-derivation},
\begin{equation}
\mathrm{Tr}\!\left[\rho_1(t)\, h_1\right]
=\big\langle\psi(t)\big|h_1\otimes\mathbb I_2\big|\psi(t)\big\rangle
\le E_{\mathrm{tot}},\qquad  \forall t\, .
\label{eq:energy-expect}
\end{equation}
Thus $\rho_1(t)$ always lies in the set of states on $\mathcal H_1$ whose mean $h_1$ energy does not exceed $E_{\mathrm{tot}}$.

At this point we invoke the Gibbs variational principle. Among all states on $\mathcal H_1$ with energy not exceeding $E_{\mathrm{tot}}$, the entropy is maximized by the Gibbs state
\begin{equation}
\gamma_\beta := \frac{e^{-\beta h_1}}{Z_1(\beta)},\qquad
Z_1(\beta)=\mathrm{Tr} e^{-\beta h_1},
\end{equation}
for the unique inverse temperature $\beta=\beta_*>0$ satisfying
\begin{equation}
    E_1(\beta_*):=-\partial_{\beta}\ln{Z_1(\beta)}|_{\beta=\beta_*}=E_{\rm tot}\,.
\end{equation}
Consequently,
\begin{equation}
S(\rho_1(t)) \le S(\gamma_{\beta_*})=\beta_* E_{\mathrm{tot}} + \ln Z_1(\beta_* ), \qquad \forall t .
\label{eq:uniform-bound-final}
\end{equation}
This establishes a uniform-in-time entropy bound whenever $Z_1(\beta)$ is finite for $\beta>0$, a condition satisfied in compact models because their spectra are discrete. 
For the two-site rotor, taking $h_1^{\rm CR}=\tfrac12 p_1^2$ yields
\begin{equation}
Z_1^{\rm CR}(\beta)=\sum_{p\in\mathbb Z} e^{-\beta p^2},
\end{equation}
a Jacobi theta function, which is finite for all $\beta>0$. By contrast, for the non-compact free-particle Hamiltonian relevant to the coupled harmonic oscillators, the partition function $Z_1^{\rm CHO}(\beta)$ diverges for all $\beta>0$, so this argument cannot yield a uniform-in-time upper bound here.

\subsection{Ensemble-based estimates for $S^{\rm CR}_{1,\rm max}$}\label{app:estimate}
While the uniform-in-time bound guarantees that the entanglement entropy remains finite, it is quite loose (see Fig.~\ref{fig:Fig7}) and,  therefore, does not provide a close estimate of $S^{\rm CR}_{1,\rm max}$. We, therefore, benchmark $S^{\rm CR}_{1,\rm max}$ against the entanglement entropies of several statistical ensembles introduced in the main text: The \emph{diagonal ensemble} (DE), the \emph{block-diagonal ensemble} (BDE), and the \emph{generalized Gibbs ensemble} (GGE).

\begin{figure}[htpb]
    \centering
    \includegraphics[width=.9\linewidth]{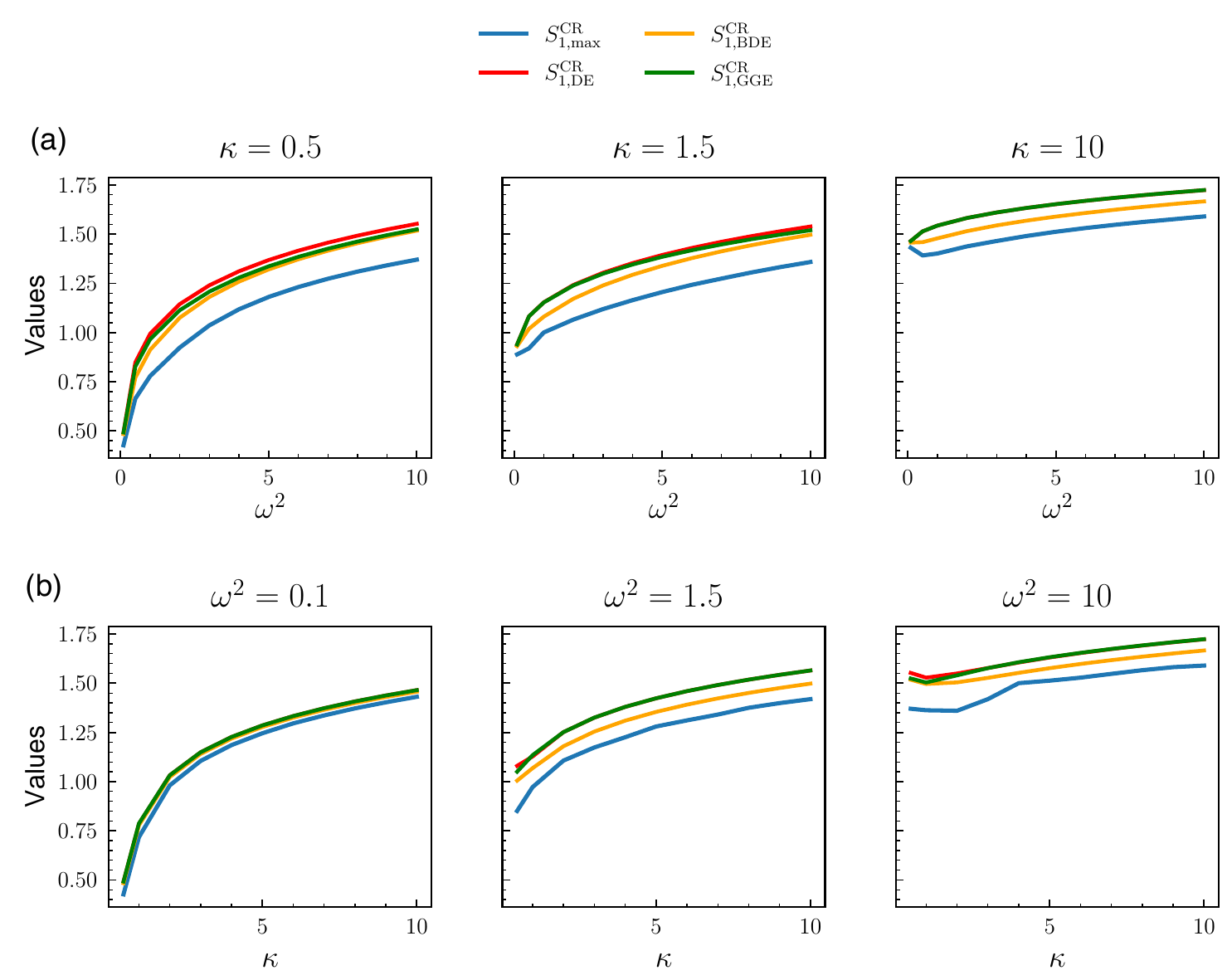}
    \caption{Ensemble-based estimates of the maximal single-site entanglement entropy in the two-site \emph{coupled-rotor} (CR) model.
Shown are the maximal entanglement entropy $S^{\rm CR}_{1,\max}$ reached during the unitary dynamics and the corresponding single-site entropies obtained from the \emph{diagonal ensemble} (DE), \emph{block-diagonal ensemble} (BDE), and \emph{generalized Gibbs ensemble} (GGE).
(a) Dependence on the initial trap strength $\omega^2$ for several fixed values of $\kappa$.
(b) Dependence on the coupling $\kappa$ for several fixed values of $\omega^2$.
Overall, the BDE provides the closest estimate to $S^{\rm CR}_{1,\max}$, with the GGE performing comparably well, especially at larger $\kappa$.}
    \label{fig:Fig8}
\end{figure}

In our numerics 
(see Fig.~\ref{fig:Fig8}), the block-diagonal ensemble yields the closest estimates to $S^{\rm CR}_{1,\rm max}$, followed by the generalized Gibbs ensemble. For large $\kappa$, the GGE value approaches the diagonal-ensemble result; for small $\kappa$ the two differ, with GGE giving a slightly better match than the diagonal ensemble.
We emphasize that one should not expect perfect agreement between these ensemble predictions and the actual $S^{\rm CR}_{1,\rm max}$ at small system sizes—this applies both to the two-body problem and to short rotor chains. As is already known from the chain of coupled harmonic oscillators (for non-critical quenches where no zero mode is generated and the entanglement indeed reaches a well-defined maximum), ensemble-based estimates only become quantitatively accurate at rather large $N$. For this reason, we expect the agreement to improve systematically as $N$ increases, although direct numerical confirmation of this trend is already unfeasible for chains as short as $N\gtrsim 4$, as discussed in Sec.~\ref{section:Sec5}.

\subsection{Benchmarking the analytic estimate of $S^{\rm CR}_{1,\rm max}$}\label{app:benchmarkGGE}
\begin{figure}[htpb]
    \centering
    \includegraphics[width=.5\linewidth]{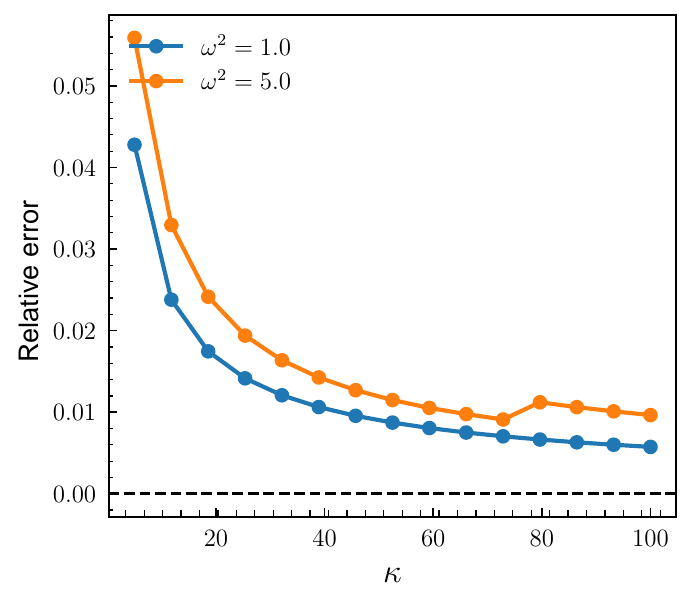}
    \caption{Accuracy of the analytical GGE estimate for the two-site \emph{coupled-rotor} (CR) model.
Shown is the relative error between the numerically obtained GGE single-site entropy $S^{\rm CR}_{1,\mathrm{GGE}}$ and the analytical approximation in Eq.~\eqref{eq:SGGE estimate}, as a function of the coupling strength $\kappa$ for different initial trap strengths $\omega^2$.
The analytical expression becomes increasingly accurate with growing $\kappa$, confirming its usefulness as an estimate for the compact entanglement ceiling.}
    \label{fig:Fig9}
\end{figure}

We benchmark the analytical approximation for the GGE entanglement entropy against the numerically evaluated GGE value in Fig.~\ref{fig:Fig9}.

\end{appendix}

\bibliography{bibliography.bib}

\end{document}